\documentclass[twocolumn]{aastex63}

\newcommand{\gaia}{{\it Gaia}}
\usepackage{xcolor}

\definecolor{my_color}{HTML}{3a18b1}

\newcommand{\teff}{\mbox{$T_{\rm eff}$}}

\newcommand{\feh}{\mbox{$\rm [Fe/H]$}}
\newcommand{\mg}{\mbox{$\rm [Mg/Fe]$}}

\newcommand{\alphafe}{\mbox{$\rm [\alpha/Fe]$}}

\newcommand{\logg}{\mbox{$\log g$}}

\usepackage{appendix}
\usepackage{amsmath}
\usepackage{mathtools}

\accepted{August 3, 2021}
\submitjournal{The Astrophysical Journal}

\shorttitle{Variance in Gaia-RVS Spectra}
\shortauthors{Rampalli et al.}

\begin{document}

\title{The Astrophysical Variance in Gaia-RVS Spectra}

\correspondingauthor{Rayna Rampalli}
\email{raynarampalli@gmail.com}

\author[0000-0001-7337-5936]{Rayna Rampalli}
\affiliation{Department of Physics and Astronomy, Dartmouth College, Hanover, NH 03755, USA}
\affiliation{Department of Astronomy, Columbia University, 550 West 120th Street, New York, NY, 10027, USA}

\author[0000-0001-5082-6693]{Melissa Ness}
\affiliation{Department of Astronomy, Columbia University, 550 West 120th Street, New York, NY, 10027, USA}
\affiliation{Center for Computational Astrophysics, Flatiron Institute, 162 Fifth Avenue, New York, NY 10010, USA}

\author[0000-0001-9116-6767]{Shola Wylie}
\affiliation{Max-Planck-Institut f{\"u}r Extraterrestrische Physik, Gie{\ss}enbachstra{\ss}e, D-85748 Garching, Germany}



\begin{abstract}
Large surveys are providing a diversity of spectroscopic observations with \gaia\ alone set to deliver millions of Ca-triplet-region spectra across the Galaxy. We aim to understand the dimensionality of the chemical abundance information in the \gaia-RVS data to inform galactic archaeology pursuits. We fit a quadratic model of four primary sources of variability, described by labels of \teff, \logg, \feh, and \alphafe, to the normalized flux of 10,802 red-clump stars from the \gaia-RVS-like ARGOS survey. We examine the residuals between ARGOS spectra and the models and find that the models capture the flux variability across $85\%$ of the wavelength region. The remaining residual variance is concentrated to the Ca-triplet features, at an amplitude up to 12\% of the normalized flux. We use principal component analysis on the residuals and find orthogonal correlations in the Ca-triplet core and wings. This variability, not captured by our model, presumably marks departures from the completeness of the 1D-LTE label description. To test the indication of low-dimensionality, we turn to abundance-space to infer how well we can predict measured [Si/H], [O/H], [Ca/H], [Ni/H], and [Al/H] abundances from the \gaia-RVS-like RAVE survey with models of \teff, \logg, \feh, and \mg. We find that we can near-entirely predict these abundances. Using high-precision APOGEE abundances, we determine that a measurement uncertainty of $<$ 0.03 dex is required to capture additional information from these elements. This indicates that a four-label model sufficiently describes chemical abundance variance for $\approx$ S/N $<$ 200 per pixel, in \gaia-RVS spectra.

\end{abstract}


\section{Introduction}\label{sec:intro}

With observations for over a billion stars, the \gaia\ mission has without doubt revolutionized our understanding of the Milky Way. The last two data releases (DR2 and EDR3; \citealt{dr2,edr3}), in particular, have furthered our understanding of how the Galaxy was formed and its subsequent evolution. The progress made to date will presumably continue in the coming years, with the third full data release (DR3) expected in 2022. Along with improved measurements, DR3 will also include spectra for over several million stars from the Radial Velocity Spectrometer (RVS). 

The RVS will observe approximately 150 million spectra in the Calcium (Ca) triplet region with a resolution of,  $R = \frac{\lambda}{d\lambda} = 11,200$ and with a signal to noise ratio (S/N) large enough that the radial velocities (RVs) of stars with a G magnitude $\lesssim 14$ can be measured depending on their spectral type \citep{recio-blanco}. Since \gaia\ surveys the entire sky, each star will have multiple observations, and the S/N will improve as these are combined \citep{dr2spec}.  
By the end of the mission, it is assumed there will be over 40 epochs of observations for most stars, with which stellar parameters such as effective temperature, surface gravity, metallicity, and $\alpha$-enhancements can be derived. In addition, depending on the quality of the combined spectra a set of individual abundances including iron, calcium, magnesium, titanium, and silicon will  be measured for $> 5$ million stars.

The Ca-triplet wavelength region in general has long been leveraged not only for RVs, but also for measurements of stellar metallicity \citep[e.g.][]{Sakari2015, Carrera2007}. Empirically, the wing-dominated Ca-triplet equivalent widths have been calibrated and validated as a precise metallicity index, which is particularly valuable at low metallicities \citep[e.g.][others]{Starkenburg2010, DaCosta1991,Waggs2019}. In addition to metallicity, the wings of the Ca-triplet  have demonstrated dependence on stellar parameters of effective temperature and surface gravity \citep[][]{Chmeilewski2000}.

The Ca-triplet region is generally inaccurately synthesised in stellar models under the one dimensional, Local Thermal Equilibrium (1D LTE) assumption \citep{Magic2013}. This is because the cores of the Ca lines are formed in the outer stellar layers, in the chromosphere, where this assumptions breaks down \citep[e.g.][]{Andretta2005}. Presumably, however, this also means that the \gaia-RVS spectra, with its many millions of measurements of Ca-triplet core strengths provides a wealth of data and a new direction for data-driven chromospheric studies \citep[e.g.][]{Cauzzi2008}. The Ca-wings, although formed in the photosphere, are also subject to dynamical time-varying astrophysical processes, in particular, subtle imprints from small-scale photospheric magnetic activity and pressure broadening \citep[][]{Leenaarts2006}.

Just as with the \gaia\ astrometry, the \gaia-RVS spectra presents us with an opportunity gain a holistic picture of stars all across both the galaxy and evolutionary state. Thus, it is timely to understand all of the information a spectrum contains, efficiently extract that information that we care about, and model out that which we do not. This question has been examined using higher resolution spectra of $R= 22,000-30,000$ (e.g. \citealt{deMijolla2021}). Furthermore, several groups have used other spectroscopic surveys (e.g. \citealt{bedell18,Ness19,Hayden21, Sharma2020, Ting2021}) to find that the majority of the chemical abundance variance in stellar spectra can be captured by only a few labels with small intrinsic information amplitudes imputed by each individual abundance.

Both the spectra itself and the chemical abundance measurements from the spectra have been leveraged to examine the dimensionality of the data (e.g. \citealt{ting12, pricejones}). Underlying many of these investigations has been the goal of testing the chemical space available for chemical tagging,  whereby stars can be traced back to their birth clusters using their abundance vectors \citep{chemicaltagging}. While achieving chemical tagging appears out of reach with current data \citep{Ness_2018, deMijolla2021}, the abundances provide powerful and subtle information for discriminating between birth environment as well as tagging the birth radii of stars (e.g. \citealt{Blancato2019,Nelson_2021}). Therefore, it is imperative that we investigate and attempt to quantify the dimensionality of stellar spectra. Here we turn our attention to \gaia-RVS-like spectra to prepare for the \gaia-RVS spectra that will be delivered in great number within the next few years.

We examine the dimensionality of \gaia-RVS spectra using 10,802 analogue spectra from the ARGOS survey \citep{Freeman_2012}.
In section \ref{sec:data}, we describe the spectra, abundances, and labels used in our work. We model the primary sources of variability in the ARGOS spectra using quadratic models with four labels. After subtracting these models from the spectra, we implement a principal component analysis on the residuals. We also examine how well such four-label models do in predicting abundance labels using the RAVE survey. We discuss how we constructed the four-label models for each ARGOS spectrum and RAVE abundance and the dimensionality reduction process in section \ref{sec:methods}. In section \ref{sec:results}, we summarize the results of the residuals from the ARGOS spectra, the results of the abundance inference with RAVE, and the principal component analysis of the ARGOS residuals. Finally, we discuss and conclude with the implications of our findings in section \ref{sec:discussion}.

\section{Data} \label{sec:data}

\subsection{ARGOS Spectra}
In this work, we primarily use spectra from the Abundances and Radial velocity Galactic Origins Survey (ARGOS). Conducted from 2008-2011, the primary intention of ARGOS was understanding the origin of the Galactic bulge \citep{Freeman_2012}. This survey obtained $\sim$28,000 red giant spectra toward the bulge and the inner disk using the AAOmega fibre-fed spectrograph \citep{sharp06} on the Anglo Australian Telescope at Siding Spring Observatory. With a resolution of R$ = 11,000$ and a wavelength region of 8400-8800 \AA\ (the Ca-triplet region), accurate RVs, stellar parameters, and $\alpha$-abundances were derived. These results helped identify the relationships between the metallicity, kinematics and orbital structure of the stars in the inner Milky Way \citep{Ness_2013}. Here, we use these spectra as a proxy for the upcoming \gaia-RVS spectra. 

To maintain uniformity for our dimensionality analysis, we select the spectra of stars in a narrow region of \logg\ across the red clump region. We do this by choosing stars with a surface gravity 2 $<$ \logg\ $<$ 3. 
From this selection we are left with 10,802 ARGOS spectra. The ARGOS spectra we work with have an average S/N = $41 \pm 11$ per pixel.

For our analysis, we are interested in extracting information from the residuals of the spectra after removing variability associated with effective temperature (\teff), metallicity (\feh), surface gravity (\logg), and $\alpha$-enrichment (\alphafe). Therefore, it is imperative to have precise labels of these variables. For this reason, we use the labels from the A2A catalog \citep{Wylie2021}, which contains 21,000 ARGOS stars with stellar parameters and abundances recalibrated to the Apache Point Observatory Galactic Evolution Experiment (APOGEE; \citealt{ravedr5,apogee}) scale. These labels were inferred with the Cannon \citep{cannon} using a training set of stars that were observed by both ARGOS and the APOGEE survey, the latter having a significantly higher resolution \citep{apogeelabels}. The precision on the labels of \teff, \logg, \feh, and \alphafe\ is  80 K, 0.19 dex, 0.1 dex, 0.07 dex, respectively.

\subsection{RAVE-on and APOGEE Abundances}

In addition to spectra, we use individual chemical abundance measurements from two spectroscopic surveys: the RAdial Velocity Experiment (RAVE) and APOGEE. The RAVE survey is an optical survey with a resolution of R = 8,000 and was used from 2003-2013 to monitor randomly chosen stars in the southern hemisphere. 
We use the RAVE-on stellar parameters and abundance labels from \citet{Casey2017}, which used The Cannon to derive more precise labels than those provided by RAVE (hence the refined moniker). APOGEE is an infrared survey with a resolution of R=22,500, designed to measure spectra for red giant stars in the inner disk and bulge. We use DR16 stellar parameters and abundance measurements from \cite{Jonsson2020}, which were provided by APOGEE's pipeline ASPCAP \citep{aspcap}. We use the 7 individual abundance measurements from the RAVE-on survey, [Fe/H], [Mg/Fe], [Si/H], [Ca/H], [O/H], [Ni/H], [Al/H], that were also measured by APOGEE. These 7 elements are of utility since \gaia-RVS will have spectra from a narrow wavelength region, and these are the elements present in that region.

\section{Methods} \label{sec:methods}

\subsection{Data Cleaning}

 Using the measured RVs from ARGOS, we RV-correct the 10,802 red clump spectra and interpolate them onto a common wavelength grid. We then continuum-normalize the spectra by dividing each by a smoothed version of itself. We create the smoothed version of each spectra using \textit{scipy.ndimage.gaussianfilter}, which uses a 1D-convolution filter to smooth a window of data. We choose a window size of $\sigma$ = 70 ($\approx 16$ \AA). Dividing by this smoothed spectra normalizes each spectra to a flux level of $\sim 1$, where absorption features are still present. Then, within each spectrum, we remove the first and last 100 \AA. This is done since a significant portion of the spectra do not extend all the way out to these wavelengths. Additionally, in those spectra that do cover the entire wavelength regime, we visually identify these regions to be noisy with flux values of zero due to bad pixels. Thus, this removal ensures that we are left with the most robust regions of the spectra on which to investigate information contained within the spectral residuals.

\subsection{Four-label Model}
The dimensionality of stellar spectra is defined by its physical properties, i.e. mass, age, evolutionary state, chemical abundances, as well as imprints and artefacts of the infrastructure used to observe it, such as the interstellar medium and conditions of the night sky. Our goal is to learn what the dimensionality of a spectrum is that underlies its physical properties once we subtract the main sources of variability that are defined by evolutionary state and overall chemical abundance. 

To build our model for each star we use the four labels, effective temperature (\teff), metallicity (\feh), alpha-enrichment\footnote{The $\alpha$-element abundance is the [Mg/Fe] abundance (propagated from the catalogue of \citealt{Wylie2021}), which we have chosen as a representative $\alpha$-element.} (\alphafe), and surface gravity (\logg). These serve as proxies for stellar mass, age, and chemical composition. Similarly to \citet{pricejones}, we construct a polynomial model described by these labels to fit to each ARGOS spectrum.

We define a vector of n = 10,802 spectra, with flux $F_n$, wavelength values $\lambda$ (where $\lambda_{\text{max}}$ = 1168), and four labels (\teff, \feh, \alphafe, \logg). Using polynomial regression from Python's \textit{sklearn}, we find the set of coefficients and intercept that best represent the relation between the flux values at each particular wavelength and the labels. The coefficients are vectors of length $\lambda$ that describe the slopes of the different parameters, their cross terms, and their squared terms. The equation to be solved can be described as follows:
\begin{equation}
  \label{eqn:mod}
  \begin{split}
    \MoveEqLeft
    \text{F}_{n_{\lambda}} = A_{\lambda}(\teff_n) + B_{\lambda}(\feh_n) + C_{\lambda}(\alphafe_n)   \\
    &+ D_{\lambda}(\logg_n) + E_{\lambda}(\teff_n)^{2} + F_{\lambda}(\teff_n\cdot\feh_n)  \\
    &+ G_{\lambda}(\teff_n\cdot\alphafe_n) + H_{\lambda}(\teff_n\cdot\logg_n)  \\
    &+ I_{\lambda}(\feh_n)^{2} + J_{\lambda}(\feh_n\cdot\alphafe_n) \\ 
    &+ K_{\lambda}(\feh_n\cdot\logg_n)  + L_{\lambda}(\alphafe_n)^{2} \\ 
    &+ M_{\lambda}(\alphafe_n\cdot\logg_n) + N_{\lambda}(\logg_n)^{2}  + O_{\lambda},
  \end{split}
\end{equation}

where $n$ is the star number, $\lambda$ is the particular wavelength, and $A_{\lambda}$, $B_{\lambda}$, $C_{\lambda}$, $D_{\lambda}$, $E_{\lambda}$, $F_{\lambda}$, $G_{\lambda}$, $H_{\lambda}$, $I_{\lambda}$, $J_{\lambda}$, $K_{\lambda}$, $L_{\lambda}$, $M_{\lambda}$, $N_{\lambda}$, and $O_{\lambda}$ are the coefficients and intercept we solve for. 

For one given star, we remove this spectrum of interest from the entire set of spectra and learn the coefficients and intercept that best represent the range of flux values found at each wavelength. These coefficients and intercept are then used to generate a model spectrum for the star of interest using the equation above, given its stellar labels. We run this procedure for every star.

We also construct a linear model for each wavelength value to compare a simpler modeling approach. We find that the differences between the linear and polynomial models and their respective residuals are on the order of 0.01\%. We use the more flexible polynomial model for this work.

\begin{figure*}
    \centering
    \includegraphics[width=7in]{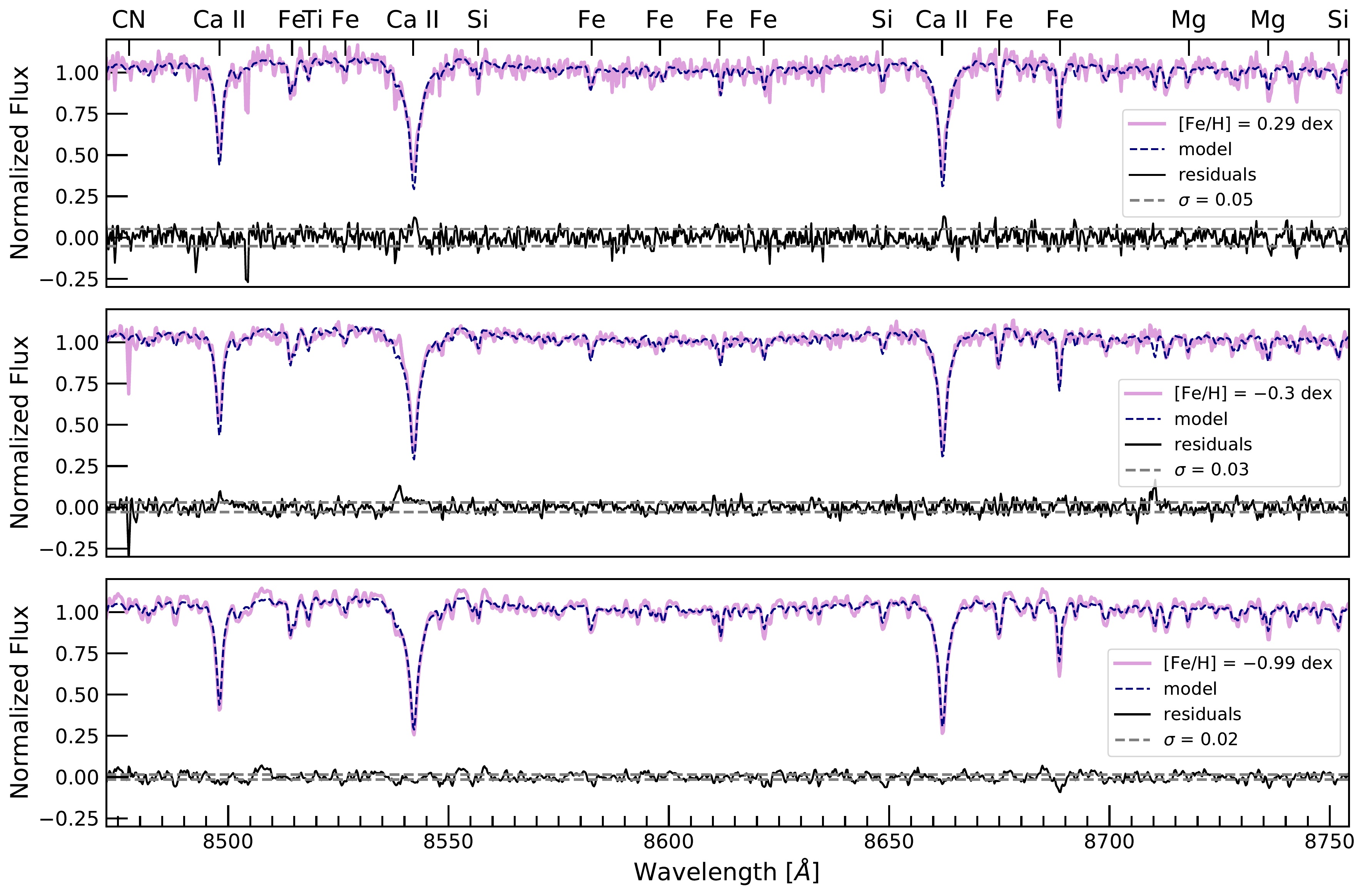}
    \caption{Examples of spectra and their four-label models with prominent absorption features labeled. Data in pink with model over-plotted with blue dashed line. Residuals in black and noise threshold plotted with grey dashed line. \textit{Top}: $[\text{Fe}/\text{H}] = 0.29$ dex. \textit{Middle}: $[\text{Fe}/\text{H}] = -0.3$ dex. \textit{Bottom}: $[\text{Fe}/\text{H}] = -0.99$ dex. }
    \label{fig:exmodspec}
\end{figure*}

Once a model is generated for each star, we then subtract the model from the spectrum to obtain a residual flux vector, $R_{n_{\lambda}}$, for every star. In Figure \ref{fig:exmodspec}, we show examples of our residual results by randomly choosing three stars from a high ($\feh > 0.2$), an average ($0.1 > \feh > -0.4$), and a low ($\feh < -0.5$) metallicity bin. For each star, we show the spectrum, the generated model, the residuals, and the noise thresholds (the square root of the mean unnormalized flux). At the top of the figure, we also show where the absorption features associated with certain elements in the spectra are located.
 
From Figure \ref{fig:exmodspec}, it is clear that the four labels capture the vast majority of the spectral variance with the residual level above the noise at the level of a few percent in only a few of the wavelength values. Systematically throughout these three stars and the rest of the sample, the most obvious region that is not entirely captured by the model is around the three Ca-triplet lines. We test that we do well in fitting out these variables with our polynomial modeling, by checking there are flat relations between the four labels and the flux of the n stars, in particular around the Ca-triplet region where we see the residual amplitude is highest. These flat relations are seen in Figure \ref{fig:corewing} in Appendix \ref{sec:caresiduals}. 

We follow the same simple regression approach with an analysis predicting abundance measurements as well. However, now instead of predicting spectra given four labels, we predict abundance labels given the same four labels. We use RAVE-on and APOGEE abundances to explore how well we can determine the 5 individual element abundances ([Si/H], [O/H], [Ca/H], [Al/H], [Ni/H]) using four labels of evolutionary state (\teff, \logg) and two abundances (\feh, \mg).

Using Equation \ref{eqn:mod}, but for a single abundance measurement, we find the coefficients and intercept to implement on stellar labels that best represent the abundances measured for the entire sample. This regression infrastructure is the same as that of our spectral analysis, except that we break up the data into a 50\% training set and 50\% test set. We respectively work with n= 150,000 APOGEE stars and n= 60,000 RAVE stars to learn a model that infers the 5 individual chemical abundances from stellar \teff, \logg, \feh, and \mg, and test how well we infer the abundances of [Si/H], [O/H], [Ca/H], [Al/H] and [Ni/H] with the remaining 50\% test set.

\subsection{Dimensionality Reduction with Principal Component Analysis}\label{sec:pcamethod}

For our residual ARGOS spectra, $R_{n_{\lambda}}$, we perform a principle component analysis (PCA). The idea being, that the features in the residuals are no longer the result of \teff, \logg, \feh, and \alphafe, but rather the result of chemical abundances, and other data artefacts. By doing the PCA, we decompose this residual information into vectors of orthogonal variance. In doing so, we hope to gain insight into underlying physical variance of the population. 

As outlined by \citet{pcaref1}, PCA reduces the sample to the number of components relevant to understanding the major variances across the sample. Mathematically speaking, PCA starts with a column space that, in this case, is defined by the number of wavelength values ($\lambda_{\rm max} = 1168$) in each spectrum. A matrix is then formed with the correlation coefficients between all the pairs of flux measurements; this is then diagonalized to find the corresponding orthogonal eigenvectors and eigenvalues. The largest eigenvalue and its associated eigenvector correspond to the direction with the largest variance and so on for the subsequent eigenvalues. These eigenvectors are the principal components (PCs). While it is hard to assign astrophysical interpretation to the PCs, we can begin to identify key features within the component and to which element they correspond. We use the \textit{sklearn} package in Python to apply this analysis.

By running the PCA on the residuals, in principle we can capture the vectors of orthogonal variations from additional abundance ratio variations since effects of \teff, \logg, \feh, and \alphafe\ have been removed. However, we also capture sources of structure that, for the purpose of our analysis, are nuisances. These include artifacts like imperfect sky subtraction and cosmic ray removal, scattered light, and instrumental imprints that may change across fiber number. Indeed, we find that a large number of PCs are needed to explain the majority of the variance in the spectra, much of which we expect arises due to these nuisance structures. We therefore do not use PCA as a tool to interpret the dimensionality of the data. Rather, PCA serves only to identify regions of the spectra with correlated variance that is not modeled out by our four labels of astrophysical variance.
PCA has been used in prior literature for similar purposes (see \citealt{pricejones,ting12}) and other areas of astronomy as well \citep{pca2}.

\begin{figure*}
    \centering
    \includegraphics[width=7in]{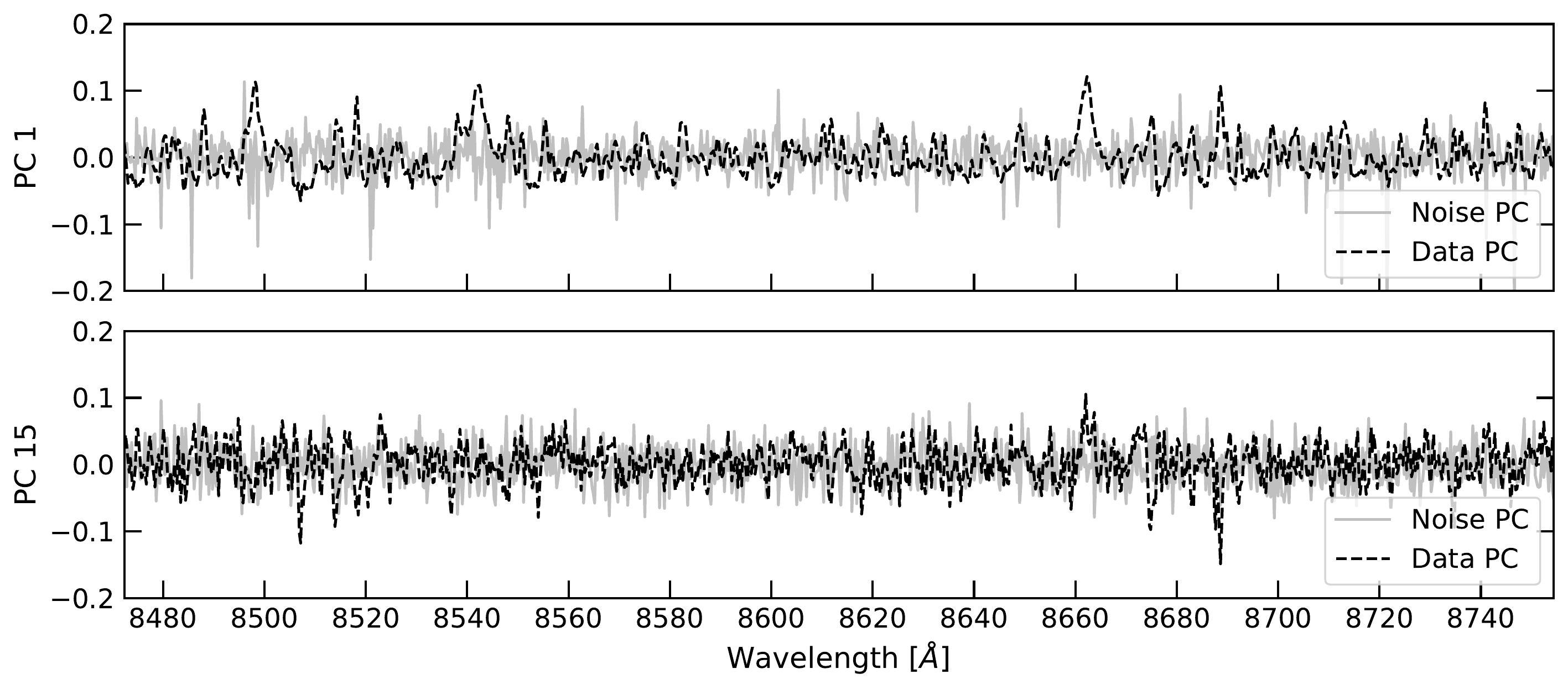}
    \caption{Noise PC versus solar-metallicity data PC for 1st and 15th PCs. \textit{Top}: First PC of noise and data. The PC from the data resembles that of a spectrum, and there are clear features as opposed to the PC generated from ``noise''. \textit{Bottom}: Fifteenth PC of noise and data. The PC from the data resembles that of the PC generated from noise indicating the PCA is capturing noise. For this reason, we choose to only examine the first four PCs.}
    \label{fig:noisepc}
\end{figure*}

On inspection of the first 15 PCs across all of our samples (see all 15 PCs solar-metallicity sample in Figure \ref{fig:allpcs} in Appendix \ref{sec:15pca}), we observed that beyond the first 4-5 components, the components look to be stochastic and noisy. We want to understand what the amplitude of noise is in the components, given our data, below which the component residuals are not meaningful. In order to establish a noise threshold, below which the component residuals are not meaningful, we performed a bootstrapping technique where we scrambled the flux values for each spectrum and ran a PCA on the sample. We compare the first PC from this noise sample and the first PC from the data to the 15th PC from the noise sample and the 15th principal component from the data in Figure \ref{fig:noisepc}. It is visually clear that there are real features in the PC from the data, but in the 15th PC, the noise resembles the data thereby validating our decision to analyze the first few PCs. In this analysis, we demarcate these ``noise thresholds'' with dashed grey lines in our PCA results (section \ref{sec:results}) as a means of highlighting the prominent features in the PCs. We construct and subtract the label-based models from the ARGOS spectra and perform a PCA on these residuals sliced in three different ways: the entire sample, solar-metallicity stars $(0.01 > \feh > -0.01 $), and metal-poor stars 
($\feh \lesssim -0.5 $). 

\subsection{Non-linear Dimensionality Reduction}
Since PCA is a linear dimensionality reduction method, we look into the non-linear method of isometric mapping (Isomap; \citealt{Tenenbaum2000}) to compare the results. While not used as often as PCA, Isomap has been used to classify stars and other astronomical objects (e.g. \citealt{isomapex}). Isomap finds the correlation in the sample by calculating the geodesic distance between pairs of points. Once these distances are calculated, a distance matrix is created and diagonalized, this determines the low-dimensional embedding of the data (or least number of dimensions needed to capture variance). The low-dimensional embedding is made up of a number of manifolds, which is analogous to the components in a PCA.

\section{Results}\label{sec:results}

\subsection{Residuals of Model-Data for ARGOS Spectra}

\subsubsection{ARGOS Residuals}\label{sec:residuals}
After subtracting the models generated from the stars' labels from the spectra themselves, we are left with residual flux vectors. These residuals represent the flux not removed by the four labels that we model. We have validated that the model fit itself is not a source of residual flux by checking there are flat correlations between the model flux and each label. Across the wavelength regime observed by ARGOS, we see on average that $84\pm 6\%$ of the residual flux measurements are within the noise floor expected for the given S/N for each spectrum. This indicates that our model has removed the vast majority of the variance in flux.

\begin{figure*}
    \centering
    \includegraphics[width=7in]{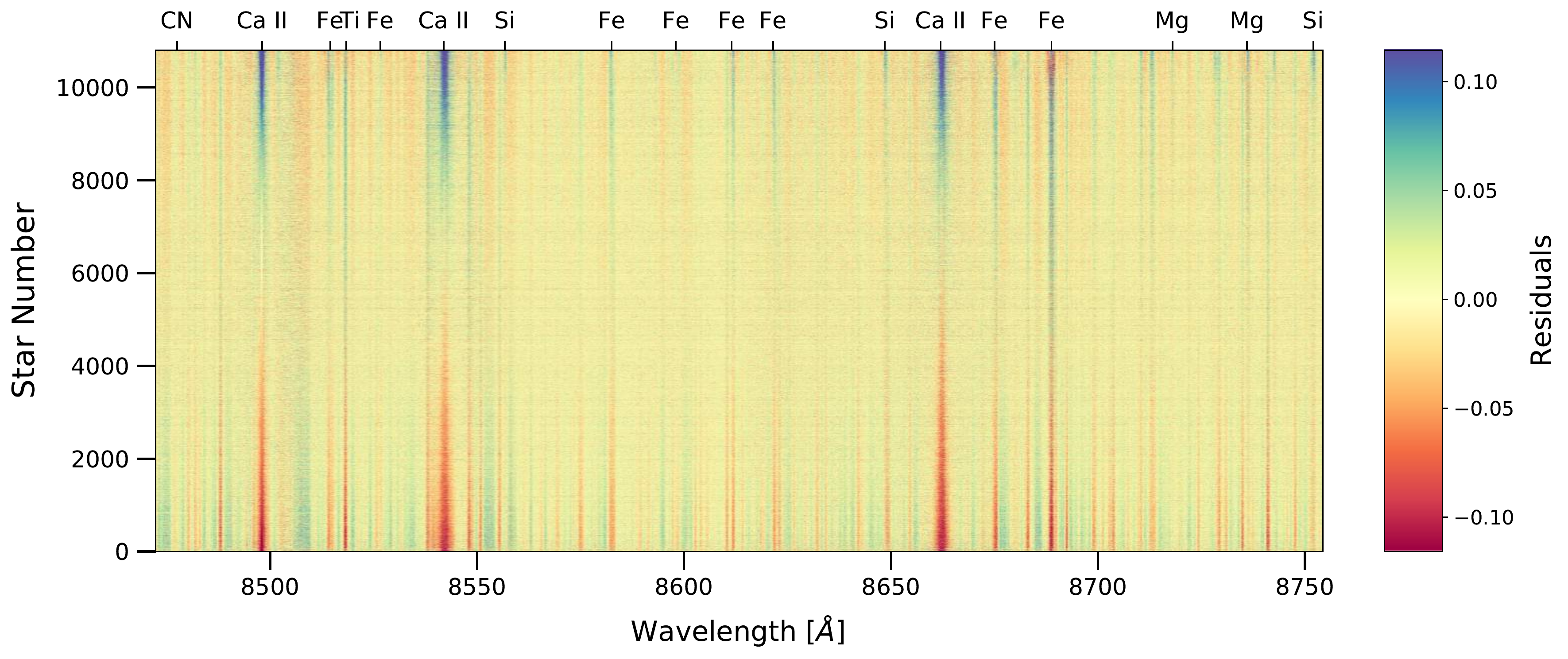}
    \caption{Residuals for each star in our sample sorted by residual value at first Ca-triplet absorption feature. For about 60\% of the stars, the models over-predict the flux values at the Ca-triplet lines. The models under-predict the flux values at the Ca-triplet lines for 40\% of the stars.}
    \label{fig:imshow}
\end{figure*}

In general the residuals with largest amplitudes are concentrated to the Ca-triplet features at 8498.7, 8542.2, 8662 \AA, respectively. These residual flux levels in the core of the triplet are on average 2\% of the normalized flux but with a significant variance of 10\%. We also find some stars with very significant residual levels in the Ca-triplet core which we discuss in Section \ref{sec:discussion}. The entirety of our samples' 10802 sets of residuals are shown in Figure \ref{fig:imshow} and are sorted by amplitude in the first Ca line. From this figure, we see that the Ca-triplet lines are 60\% overpredicted and 40\% underpredicted by the models. While the S/N at the Ca-triplet is on average 50\% lower than across the entire spectrum, we find that 83\% of the residuals around these lines are above the expected residuals from the noise.

In addition to the Ca-triplet, there are other systematic regions of residuals at the sub few-percentage level where the model under or over-predicts the data. These are usually around the strong elemental lines such as Mg, Ti, Si, and Fe. We investigate the residuals around the strongest Fe line at $ \sim8685$ \AA\ as a test case, where we find average residual flux levels of $6 \pm 5\%$. The Fe line is on average 23\% lower than the average S/N across the spectrum, but we find that 78\% of the residuals are above the expected residuals from noise. This suggests that the labels cannot precisely model the strong lines, such as Fe \citep{Holzreuter2013}, possibly a result of their derivation from 1D-LTE models.

\begin{figure}
    \centering
    \includegraphics[scale=0.53]{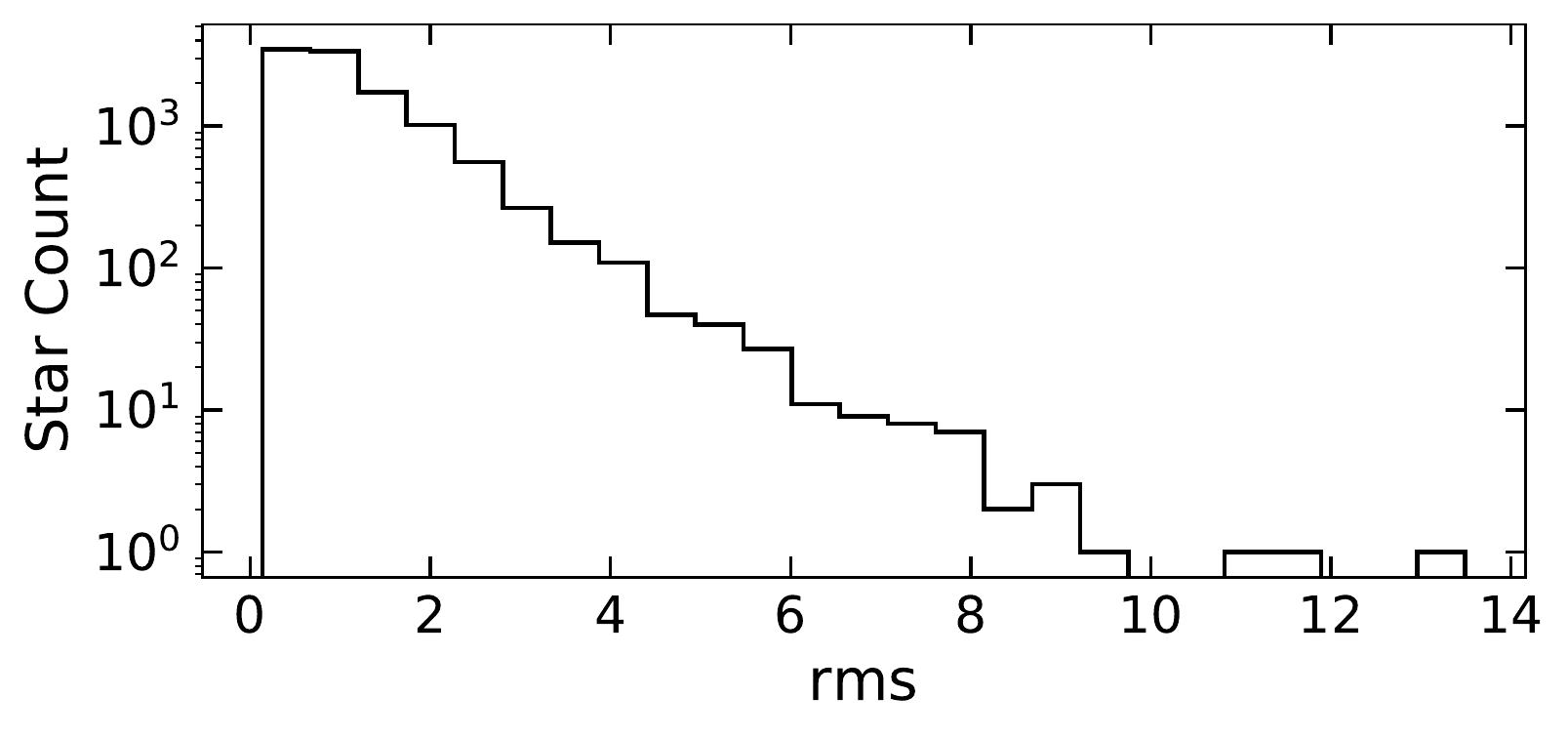}
    \caption{The rms distribution for the set of 10,802 ARGOS residuals. We calculate the rms value between the model and data. Stars with high rms ($> 5$) values either i) have spectra with bad pixels that made it through the initial data cleaning or ii) have features that are so unique compared to the rest of the sample that the model fit is poor.}
    \label{fig:chi2}
\end{figure}

We also calculate a root mean squared (rms) value for each star:
\begin{equation}
    \text{rms} = \Sigma\left( F_{\lambda}-M_{\lambda}\right)^{2},
\end{equation}

where $F_{\lambda}$ is the measured flux value and  $M_{\lambda}$ is the model value at the nth wavelength. The results are shown in Figure \ref{fig:chi2}. We choose to use rms as a metric\footnote{We use rms instead of a  $\chi^{2}$ value calculation, as our error estimates at each wavelength are based only on $F_{\lambda}^{1/2}$, and we do not have a mask to consider bad pixels, cosmic rays or imperfectly subtracted sky.}. We inspect 107 stars with an rms $> 5$ and find that most have spectra with bad pixels or a low S/N. However, we also find that about 15\% of these spectra have strong features that appear astrophysical rather than a result of systematics. This is mostly caused by the anomalous behaviour at the Ca-triplet lines. We have noted some of these outliers in Figure \ref{fig:outliers} and describe them further in section \ref{sec:outliers}. 

\subsubsection{Significance of Label Errors}
In order to ascertain the validity of our results, we consider the significance of the label errors on our models and subsequent residuals. 
We perform a Monte Carlo sampling test. For each star and each label, we draw 75 samples from a Gaussian distribution with a mean equal to the original label value and a standard deviation equal to the label uncertainty. With these new labels, we construct a new set of spectral models and corresponding residuals. We find that the standard deviation of the 75 sets of residuals for each star and label is so small that it is negligible, thereby indicating the label errors do not affect our results.

\subsubsection{Significance of Label Precision}
\begin{figure*}
    \centering
    \includegraphics[scale=0.5]{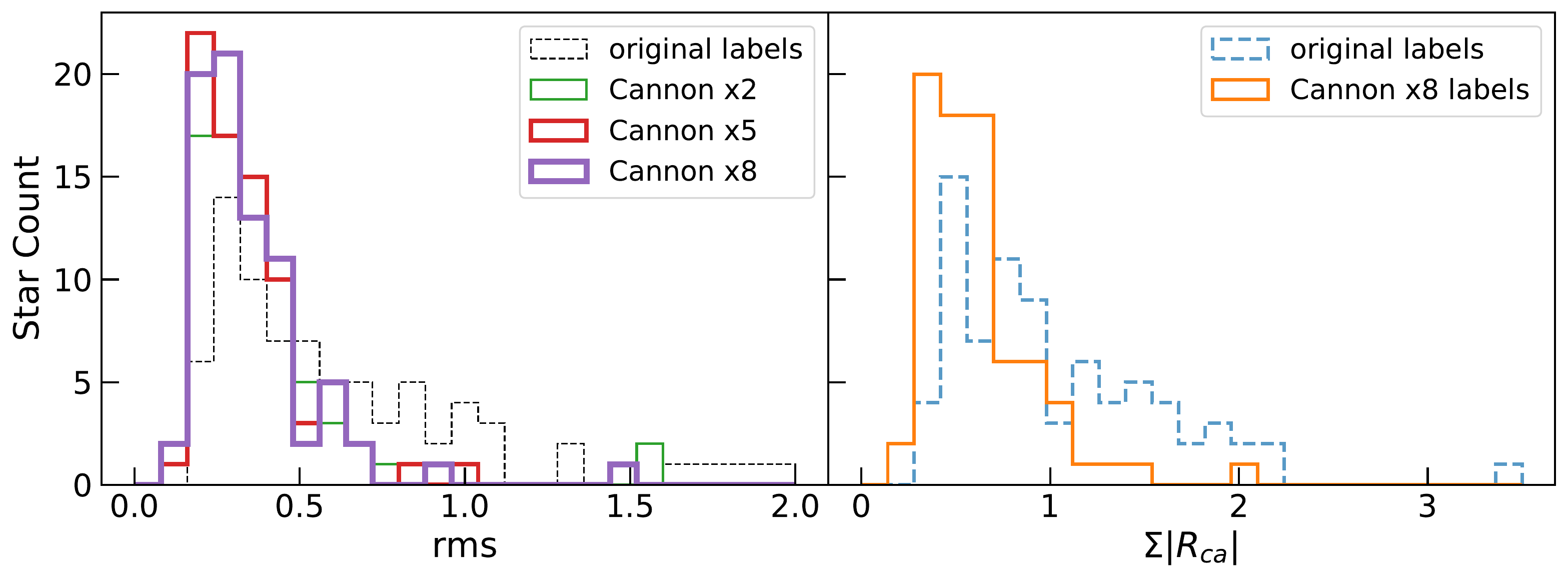}
    \caption{Tests of label precision in calculating the residuals. \textit{Left}: rms distribution of model fits after inferring new labels with the Cannon. Using a sample of 75 high S/N stars and their original labels, we use the Cannon to infer new sets of labels 8 times. After each iteration, we calculate new models and residuals and report the rms value. We note a 30\% decrease in residuals by the 8th iteration. \textit{Right}: Distribution of the sum of the residuals around a Ca-triplet line. With the same 75 stars, we calculate the amplitude of the residuals around one of the Ca-triplet lines and compare the results from the original labels to the more precise Cannon-inferred labels. We find that 25\% of the stars fall outside of the standard deviation of the Cannon-inferred label distribution, suggesting that this is the group of stars driving the decrease in the residuals seen in the left panel.}
    \label{fig:ltests}
\end{figure*}

We also consider the precision of the labels by running a test with the Cannon \citep{cannon}. Using a sample of 75 stars with highest S/N, we use the original labels as our training set and have the Cannon infer a new set of labels with which we construct new model spectra and subtract from the ARGOS spectra to get new residuals. We iterate through this process 10 times, using the previous iteration's inferred labels as the training set from which to infer the current iteration's labels. We find that as the Cannon infers new labels, the new models' residuals shrink. The largest decrease we see is by about 30\% in the 8th iteration after which the decrease in residuals hits a ceiling. We show the rms after each iteration on the left hand panel of Figure \ref{fig:ltests}.

To understand the effect of using more precise labels, we look at the sum of the amplitude of the residuals around one of the Ca-triplet lines in the right hand panel of Figure \ref{fig:ltests}. Twenty-five percent of the stars have residuals from the original labels that are larger than the standard deviation of the residuals from  the Cannon-inferred labels. This suggests that a quarter of the stars are driving the decrease in residuals discussed above. The reason we do not see this effect in our test of label errors is because the Cannon has inferred more a precise label description of the flux, which are outside of the error assigned to the original labels.

\subsubsection{Inferring Abundances}\label{sec:ia}

\begin{figure*}
  \centering
  \includegraphics[angle=270,width=1.0\textwidth,trim={7cm 1.5cm 7cm 2cm},clip]{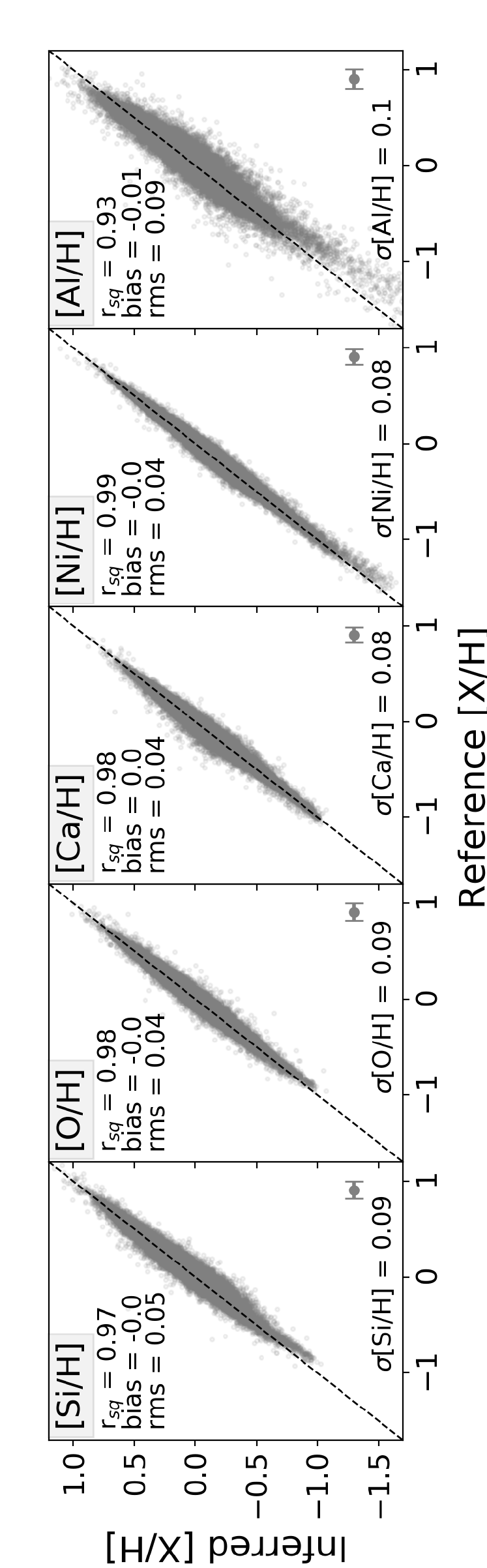}
    \includegraphics[angle=270,width=1.0\textwidth,trim={7cm 1.5cm 7cm 2cm},clip]{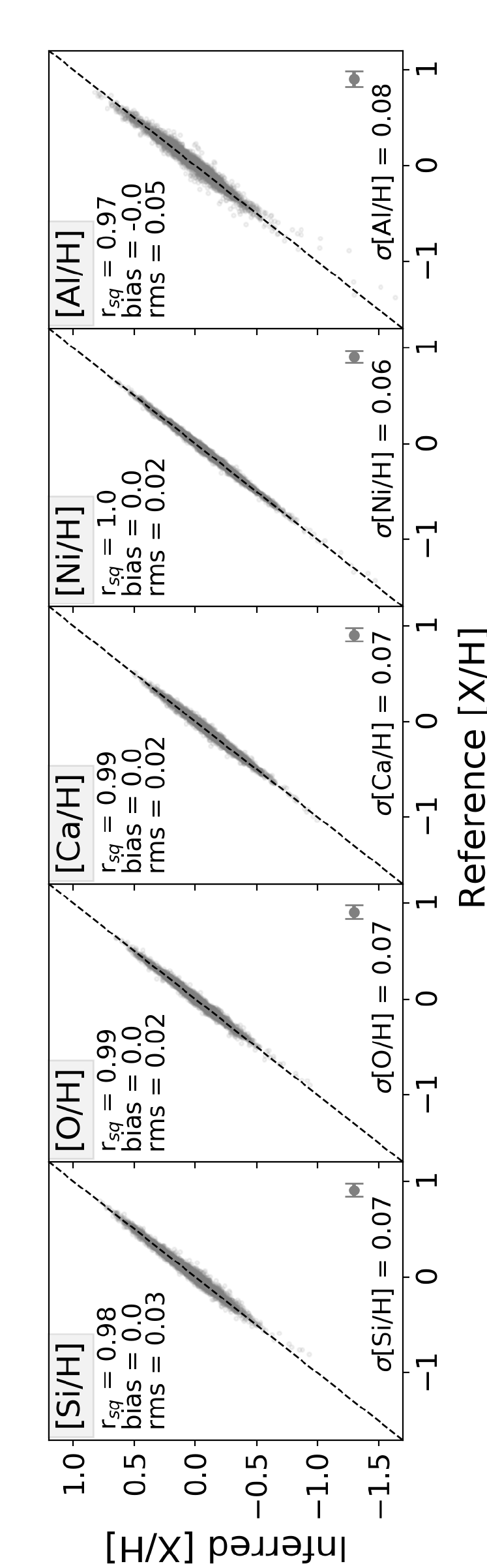}
      \caption{The results for the test set of RAVE-on stars (across -1 $<$ \logg $<$ 4 dex from \citealt{Casey2017}). Each panel shows the r-squared goodness of fit metric of the model, the bias and the rms difference of the (Reference-Inferred) abundance. The mean error is also included at bottom. The prediction is to within (or better) the error reported on the reference individual abundance labels, meaning four labels are sufficient to describe the full set of abundances to this precision. \textit{Top}: Results for all RAVE-on 50,000 stars. \textit{Bottom}: Results for the 3000 test stars with S/N per resolution element $>$ 80 (S/N = $80-120$) showing the rms uncertainty decreases further at high S/N.}
    \label{fig:elem}
\end{figure*}

Although our analysis of the spectra indicates that most of the variance (84\%, see Section \ref{sec:residuals}) is captured by the four labels we use in our models, there are other transitions in this wavelength region including the elements Al, Ni, O, Si, and Ca. Thus we expect that we should be able to infer the other chemical abundances that are present in this region using the four labels alone of \teff, \logg, \feh, and \mg\ at comparable S/N. We test this proposition using the RAVE-on catalogue, which contains 150,000 abundances across a range of stellar evolutionary states (from $-1 < \text{\logg} < 4$). The S/N of these stars is $\approx$ 45 $\pm$ 16 per pixel.  We use a second order polynomial regression (using python's \textit{sklearn} library, as outlined in Section \ref{sec:methods}) to predict the individual abundances\footnote{Similar to with the ARGOS data, we note that the model performs only marginally worse with linear regression rather than with the our second-order polynomial model. The element Al shows a small bias at low metallicity, but accounting for this bias, its prediction would be unhindered.}. We divide the data up into a random subset of 50,000 training objects and 50,000 separate test objects. We predict the 5 chemical abundances reported in RAVE-on of [Si/H], [Ca/H], [O/H], [Ni/H], [Al/H]. The prediction of these elements is shown in the top panel of Figure \ref{fig:elem}.

Figure \ref{fig:elem} reports the element in each sub-figure at top left in each sub-panel. The bias and rms difference of the predicted-reference abundance is indicated, as is the model's R-squared (r$_{sq}$) value. This measures the model performance using the training set of stars and is the proportion of the variance of the predicted value that is explained by a reference value (a value of 1 is the highest score). The typical systematic uncertainty of each measurement is reported in the right hand corner of each sub-panel. In the bottom panel of Figure \ref{fig:elem} we show the subset of stars with S/N $>$ 80 to highlight how tight the relations are for the high quality spectra.

That the $\alpha$-elements (Ca, Si, Mg) are correlated is unsurprising, given their common (mostly) type II supernovae source. However, our results shown in Figure \ref{fig:elem} demonstrate that we can predict all five chemical elements to the precision with which they are reported using the four labels of \teff, \logg, \feh, and \mg\ alone. 
In fact, we can predict these elements to a higher precision then their reported systematic uncertainties. We suspect this is because their errors are correlated. We note that the prediction of Ca from Mg is indicative that the residuals that we see around the Ca-triplet in our spectra when we subtract our four label model from our data are not imputed by Ca itself, but are instead from some other astrophysical mechanisms as we discuss further in section \ref{sec:PCA}.

\begin{figure*}
  \centering
  \includegraphics[angle=270,width=1.0\textwidth,trim={7cm 1cm 7cm 2cm},clip]{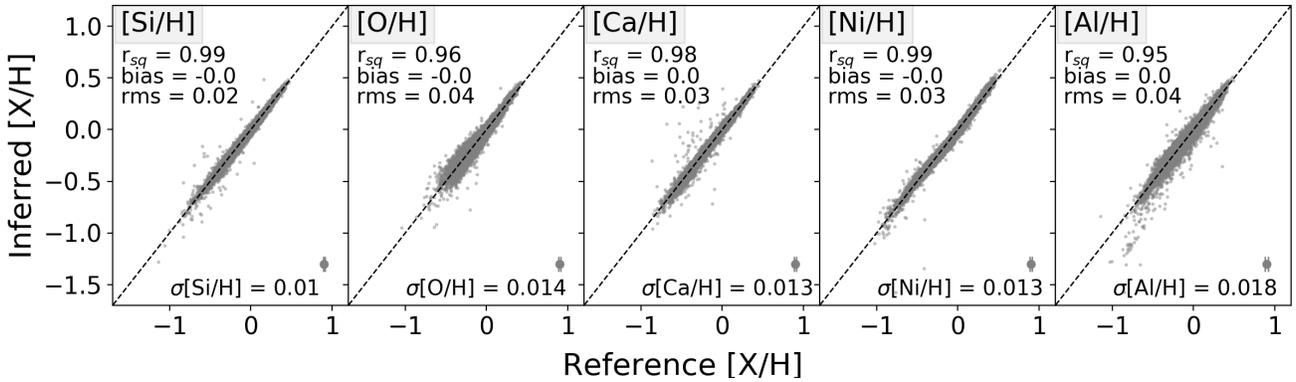}
      \caption{As per Figure \ref{fig:elem} but for  $\approx$ 20,000 APOGEE red giant stars with S/N $>$ 200. The mean intrinsic dispersion of these elements once the four labels of \teff, \logg, [Fe/H] and [Mg/H] have been accounted for is 0.03 dex (and above the measurement precision of APOGEE for these elements).} 
    \label{fig:elemap}
\end{figure*}

To demonstrate that this ability to learn abundances from a four-label model alone is a generic result of abundances, and not specific to the wavelength, methodology or resolution, we also perform a similar test using APOGEE data with the same subset of elements. As was done with the RAVE-on abundances, we divide the data into a training and test set. This time, of $\approx$ 20,000 stars each. The results at test time of training a model to predict the abundances using the four-label model for red giant stars with $0 < \text{\logg} < 3.5$ dex and $4500 < \text{\teff} < 5500$ K is shown in Figure \ref{fig:elemap}, for stars with S/N $>$ 200. The difference between the APOGEE and the RAVE-on results is that the APOGEE rms difference between the reference and inferred abundance for each element is higher than the uncertainty on the measurements. The intrinsic information in the abundances once \feh\ and \mg\ are accounted for is 0.02 dex for [Si/H], 0.03 dex for [Ca/H] and [Ni/H], and 0.04 dex for [O/H] and [Al/H]. Thus, these elements need to be measured to a precision of, on average, $<$ 0.03 dex to get to the intrinsic information in them beyond that captured in \feh\ and \mg. This is important to consider since \gaia-RVS will observe a subset of spectra for a star that can be co-added such that it is sufficient S/N to reach this precision which will allow for measuring abundances down to this level of precision (further discussed in section \ref{sec:discussion}).

These regression results are also in effect a recasting of the result in which the intrinsic dispersion of elements around their age-abundance relations are small \citep[e.g.][]{bedell18, Ness19, Sharma2020, Hayden21}. Therefore, our results here are not necessarily surprising. Furthermore, the recent work of \citet{Ting2021} has similarly demonstrated the same small intrinsic dispersions in additional elements once the information in a few is accounted for in the APOGEE spectra. 

These results demonstrate that while the spectra are fairly low-dimensional overall, there are still important details that can be captured. In surveys like APOGEE that can currently measure abundances to precisions of $\approx$ 0.03 dex or better, it is clear that there is intrinsic information in each abundance not captured in for example the four-label model, and in subsequently, in element correlations of these residuals. While the abundance measurements of lower-resolution surveys like RAVE and ARGOS do not offer intrinsic information, at the SNR examined here, there are places in the spectra with systematic residuals; namely, around the Ca-triplet, that are worthy of understanding in detail.

\subsection{PCA}\label{sec:PCA}

\begin{figure*}
    \centering
    \includegraphics[angle=90,width=\textwidth,height=\textheight]{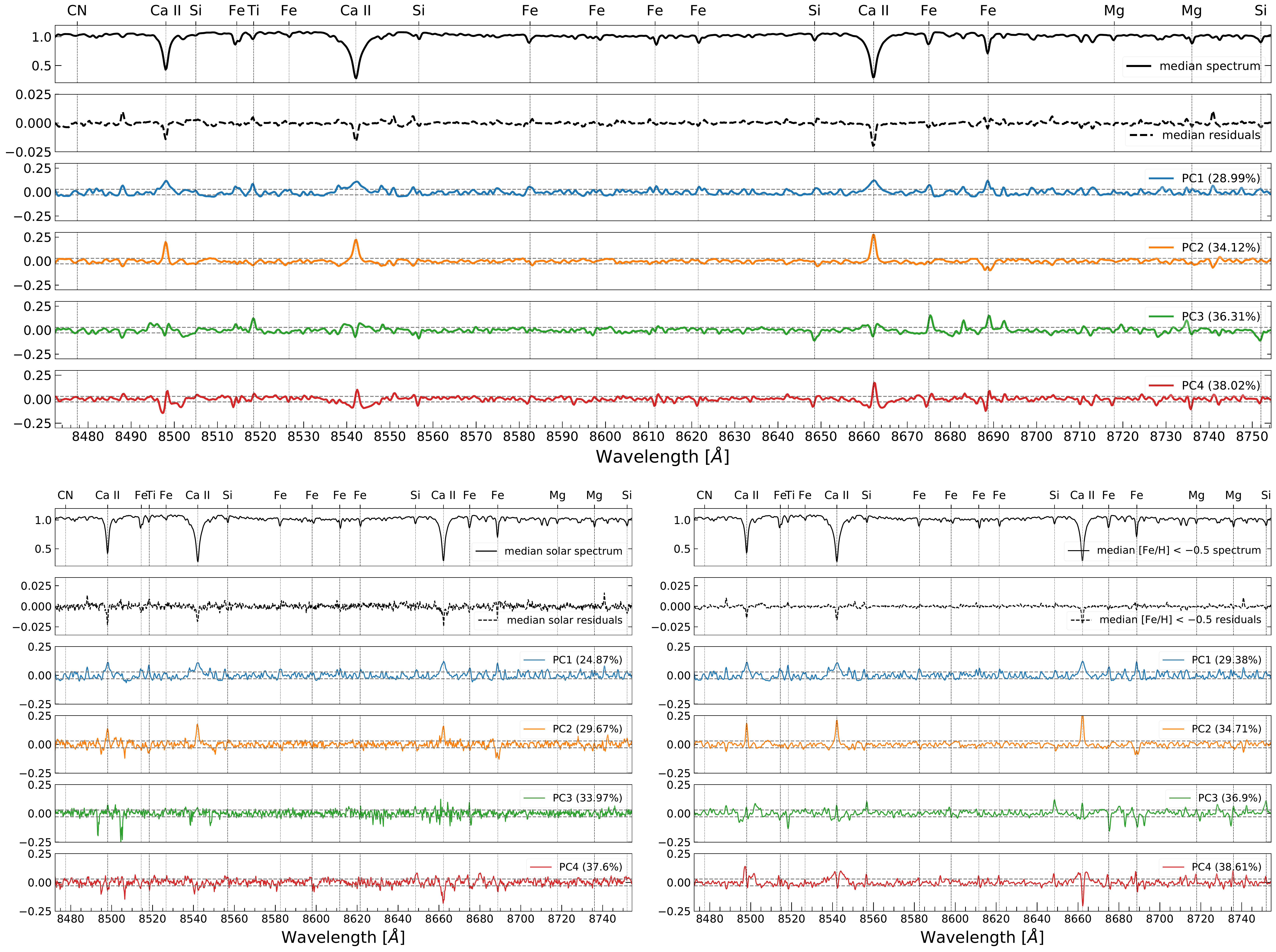}
    \caption{Median spectrum, median residuals, and first four PCs for the spectra with cumulative percentage of variance captured and noise thresholds in dashed grey lines as discussed in section \ref{sec:methods} noted. \textit{Top}: Entire sample. \textit{Bottom-left}: Solar metallicity stars ($ -0.01 \lesssim \feh \lesssim 0.01 $). \textit{Bottom-right}: Metal-poor stars ($\feh \lesssim -0.5 $).}
    \label{fig:pcaplot}
\end{figure*}

We present the first four PCs of the full set stars in our sample, the 140 solar metallicity stars ($ -0.01 \lesssim \feh \lesssim 0.01 $), and the 2805 metal poor stars ($\feh \lesssim -0.5 $) in Figure \ref{fig:pcaplot}. Since we use a PCA directly on the spectra and the number of dimensions needed to describe a majority of the spectra is high, there are many extraneous sources in our signal. These include any instrument variance over wavelength and between fibers, imperfect sky and telluric subtraction, cosmic ray imprints, and noisy spectrograph pixels, and imperfect RV corrections. Thus `unsupervised' PCA is not really a useful tool for precisely assessing the dimensionality of chemical and evolutionary state driven sources of variability in the spectra. However, the first four PCs help identify regions of the spectra that are not captured by our four-label model. 

In examining the PCA components, we aim to understand the sample holistically and what excess variability exists once the influence of four labels are removed. We focus on the most prominent features with signal above the noise thresholds as discussed in section \ref{sec:pcamethod}. 

Considering only the first four PCs, across the entire sample (top panel, Figure \ref{fig:pcaplot}), the first principal component captures 29\% of the variance in the data, the second captures another 5\% (cumulatively 34\%), the third another 2\% (36\%), and the fourth another 2\% (38\%). 

In PC 1, we see three features associated with the Ca-triplet with positive amplitudes, appearing $4\sigma$ above the noise. This is not entirely surprising since these wavelengths have the highest residuals. In PC 2, we again see features associated with the Ca-triplet, but the features are $\sim 30-50\%$ narrower compared to those seen in PC 1. They are concentrated to the cores and appear $\gtrsim 7\sigma$ above the noise. It is worth noting that the cores and wings are represented in separate PCs and are thus construed as orthogonal, indicating that two different astrophysical mechanisms account for these features. This makes sense given what we know about the cores of the Ca-triplet being associated with the chromosphere and the wings with the photosphere \citep{Andretta2005,DaCosta1991}. These results also demonstrate the need for using the spectra directly to learn what the labels and respective models do not capture.

In PC 3, we see that all three features around the Ca-triplet reverse sign and are much smaller in amplitude ($\lesssim 1\sigma$) than the previous PCs. In PC 4, we see the first Ca-triplet feature resembles a p-cygni function, while the other two resemble the Ca-triplet features in PC 2, but at smaller amplitudes. It is less obvious what these features indicate, but possible explanations include microturbulence and/or imperfect RV corrections. We also note that all the features seen across the four PCs are asymmetric about the wavelength associated with an element absorption feature. 

In addition to the Ca-triplet, we also see features in the PCA around Fe lines, although these are only at $\lesssim 50\%$ of the level of the Ca-triplet. In general, we see the PCs are very similar across the three metallicity bins.

\subsubsection{Isomap}
\begin{figure}
    \centering
    \includegraphics[scale=0.36]{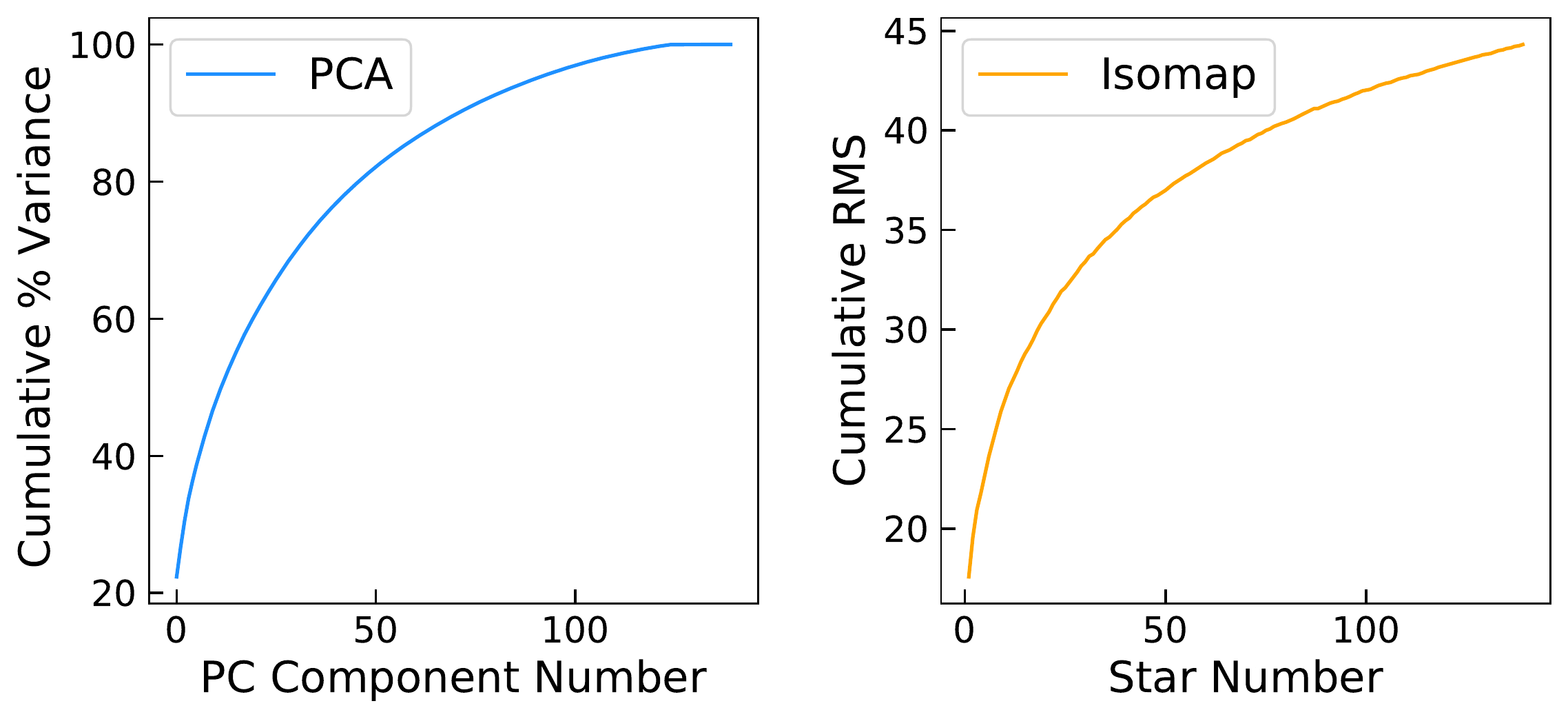}
    \caption{Comparison of PCA and Isomap. \textit{Left}: Variance captured by each PCA component in the solar metallicty sample. \textit{Right}: Cumulative rms captured for each isomap iteration through the solar metallitiy sample. Both methods have a metric that follows the same trend of flattening out over the sample.}
    \label{fig:isomapcomp}
\end{figure}

We also examined the dimensionality of the spectra using Isomap \citep{Tenenbaum2000}. Using the isomap module in \textit{sklearn}, the spectra can be represented as a series of manifolds for which geodesic distances are calculated and are then transformed into flux vectors. To compare the PCA results with isomap, as we iterate through each manifold, we calculate the rms value between the isomap transform for the current flux vector output and the prior flux vector output, and add these cumulative results. We find that this distribution resembles the cumulative variance captured by the PCA in Figure \ref{fig:isomapcomp} and does not add any new insight.  

\section{Discussion \& Conclusions}\label{sec:discussion}

Using measurements of \teff, \logg, \feh, \alphafe, and a simple polynomial regression, we fit for the variance in normalised flux across $\approx 11000$ red clump ARGOS spectra in the Ca-triplet wavelength region. We use our regression model to generate, for each star, a model flux as produced by these four parameters. We then subtract each star's model flux from its true flux to obtain a residual flux vector. On average, we find that our four-label model accounts for $> 83 \pm 5 \%$ of the variation in the flux wavelength values such that the residual flux is within the noise. In the remaining wavelengths, the amplitudes are highest from 8495.3-8501.1, 8536.4-8548.5, 8657.2-8668 \AA\ around the  Ca-triplet lines (8498.7, 8542.2, 8662 \AA). In the mean residuals alone, we see the mean amplitudes are  2\% of the flux with a variance of 10\%. Recall that in spite of the lower S/N at the Ca-triplet, we find that 83\% of these residuals are above the expected residuals from the noise. We note cases in which the labels do a poor job of modeling in section \ref{sec:outliers}.

Employing a PCA gives us the additional insight of revealing the orthogonal structure in the residuals that we see at the Ca-triplet lines. In principle, the PCA can also quantify the dimensionality of the data. However, our spectra is imputed with numerous systematics and stochastics, which makes the dimensionality high. Therefore, we use the first several components to focus on regions of the spectra not captured by the four-label model, primarily around the Ca-triplet. In the third and fourth PCs, we see features associated with the triplet at smaller amplitudes as compared to the first two components ($1\sigma$ versus $>4\sigma$). Some of these features resemble p-cygni functions and perhaps could be attributed to imperfect RV-corrections and microturbulence. The orthogonal variance in the cores and wings of the Ca-triplet seen in PCs 1-2 suggests that there are astrophysical mechanisms at work that we have not accounted for in our model. This is corroborated by the literature, which discusses how the cores and wings of the Ca-triplet are formed in the chromosphere and photosphere respectively \citep{Andretta2005,Leenaarts2006}.


The Ca-triplet residuals indicate that our parameterization of stars using static labels is, in detail, incomplete. Stars vary over time \citep{Aerts2021}, and the residuals around the Ca-triplet lines and their different distributions in orthogonal components in the PCA are likely a reflection of this. We note that the four labels in our model come from the 1D LTE framework as they are effectively derived from physical models with these assumptions. The 1D LTE set of labels is an inadequate parameterisation when the dynamical nature of the star matters in imputing the line flux, which is a possible explanation behind the high residuals at the strong Fe line at 8685 \AA\ as well \citep{Holzreuter2013}. Figure 7 from \cite{Magic2013} shows the fine granulation pattern from 3D models which give rise to turbulent convective motions in the outer layers. Due to the mechanisms of its formation, while Ca-triplet itself in its equivalent width (dominated by the wings) is primarily an [Fe/H] indicator \citep{Sakari2015,Carrera2007,Starkenburg2010,DaCosta1991,Waggs2019}, this is not in contraction with the amplitude of these residuals. We report extra variance that is largest in the Ca-triplet core at the level of a few percent, so this is a very subtle signature in the data. High resolution studies of the Sun show there are clear variations in the Ca-triplet region, including the cores and wings, due to the dynamical nature of the stellar surface \citep{Andretta2005,Leenaarts2006,Chmeilewski2000}. Based on our preliminary results, it is not hard to imagine that \gaia\ will provide a low-resolution but macro view of this astrophysical behaviour across stellar dynamical states.

Outside of the Ca lines, the remarkable fit of the four-label model to the data at the stars' S/N motivated us to examine how well we can predict individual abundances in this spectra region. We make the prediction of individual abundances with a model built on the four labels of \teff, \logg, \feh\ and [Mg/Fe] from the RAVE survey. We show in section \ref{sec:ia} that we can predict five elements, Si, O, Ca, Al, and Ni, in this spectral region to within their measurement uncertainties. Then, using APOGEE we quantified the level of precision at which these elements measure intrinsic variance not captured in \feh\ and \mg\ alone. On average this is $\approx$ 0.03 dex. From Figure 3 in \citet{Ting2017}, we see that a precision of 0.03 dex is reached at \gaia-RVS resolution of around 10,000 at S/N $\gtrsim$ 200. This precision is comparable to a result from \cite{Ting2021}, whereby after conditioning on \feh\ and [Mg/H], they predict the elements that we study with a  residual scatter of $\approx$ 0.02 dex. While this is beyond the scope of our work, \cite{Ting2021} highlight that this residual amplitude is a measure of the individual story that each element has to tell. This is confirmation that \feh\ and \mg\ do not contain all of the information in a stellar spectrum as the residual abundances are correlated between elements above observational uncertainties. 

We note that our results are for a very narrow range in stellar parameters, solely along the red giant branch. There might be a larger residual variance in the data along a larger range in parameter and evolutionary space. Nevertheless, from the test of RAVE abundance prediction, which was done across a wide evolutionary state, we found we can predict the abundances to their precision with our four label model of \teff, \logg, \feh, and \alphafe.

\begin{figure*}
    \centering
    \includegraphics[width=7in]{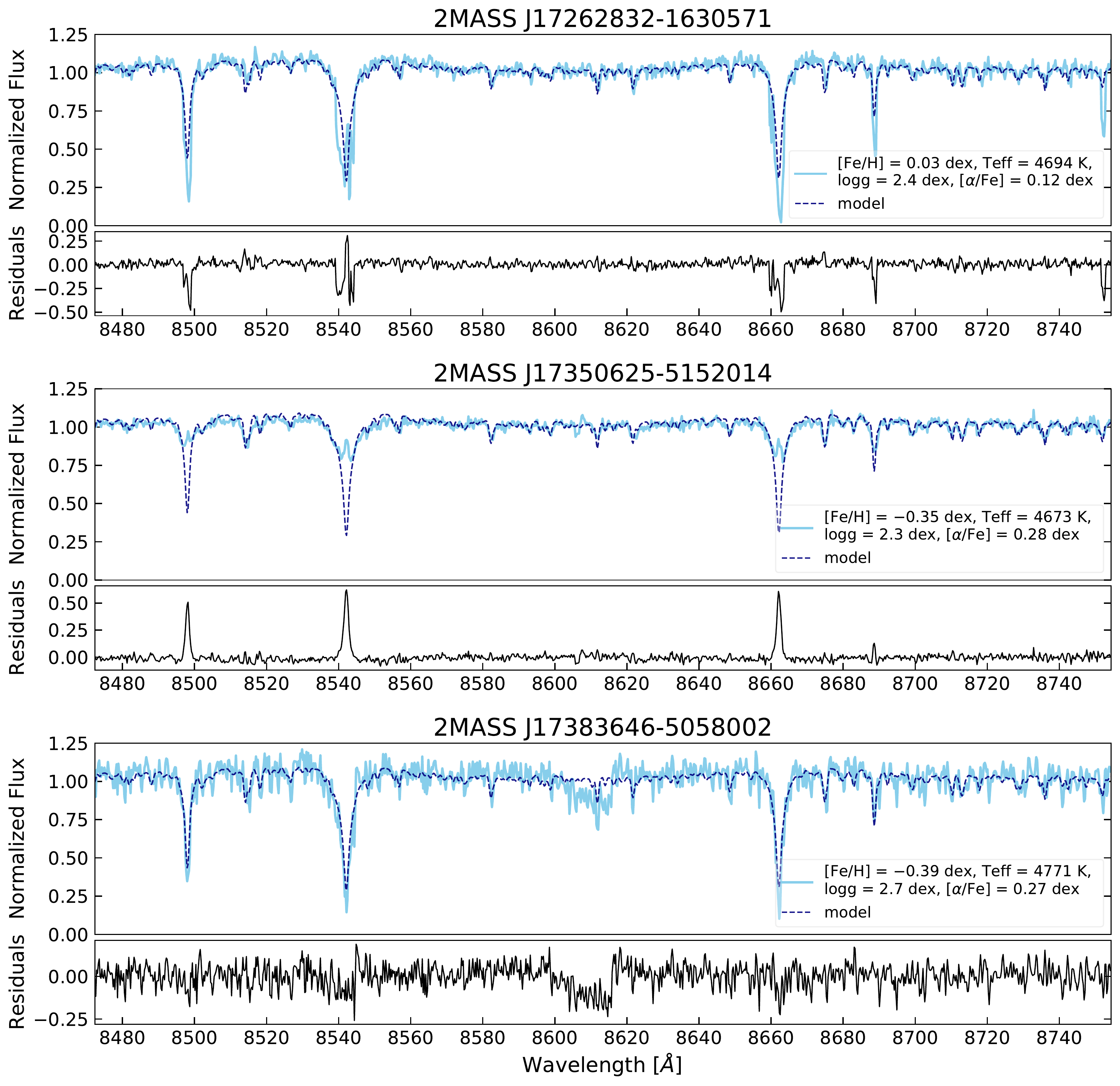}
    \caption{Select stars with poor model fits. Data in light blue. Model in dark blue dashed line. Residuals in black. \textit{Top}: Likely binary or trinary system. \textit{Middle}: Likely emission AGB star.
    \textit{Bottom}: Likely star with a molecular band around 8600-8620 \AA.}
    \label{fig:outliers}
\end{figure*}

\subsection{Outliers}\label{sec:outliers}
As shown in Figure \ref{fig:chi2}, we calculate the rms of every star and visually inspect those with an rms $ > 5$. While we identify stars that have low S/N or a bad pixel resulting in a poor fit, we also find stars with unique looking spectra illustrating the utility of this method of model-fitting for finding outliers. 
In Figure \ref{fig:outliers}, we note three such examples.

In the top panel, for 2MASS J17262832-16303571, we see what is likely a binary or trinary system given the presence of several absorption features around the Ca-triplet regions. No additional literature exist for this star that validates this finding. 

In the middle panel, for 2MASS J17350625-5152014, we see emission at the Ca-triplet suggesting it is an asymptotic giant branch (AGB) star rather than a red clump star. While this is a rarer occurrence, \citet{gomes21} discusses a scenario in which intermediate stars are rapidly rotating in the main sequence phase and have yet to spin down due to angular momentum loss. Since they do not have significant convective envelopes, the activity levels are not high. However, as they enter the giant phase, they form a convective envelope, and the rapid rotation causes high levels of activity (e.g. Ca emission). Such stars with high chromospheric activity in the same \teff\ and \feh\ ranges have also been found in the RAVE survey \citep{Zerjal13}.

In the bottom panel, for 2MASS J17383646-5058002, we see an interesting feature between 8595-8620 \AA\ that the model cannot capture. While we are unsure about what exactly this feature is, we posit that it is the result of the molecular bands.  

\subsection{Future Prospects} 
While we find that PCA is not particularly useful for understanding dimensionality of the spectra in terms of variance that links to chemical element abundances, in this work, it emphasizes how the Ca-triplet cannot entirely be explained by stellar 1D LTE models. The \gaia\ RVS spectra will present us with the opportunity to understand how the wings and core strength of the Ca-triplet can change with evolutionary state, as a result of stellar binarity, and even between observation epochs. Additionally, since we know parts of the Ca-triplet are formed in the chromosphere, it could be argued these features are also a result of magnetic activity. Magnetic activity is often quantified using X-ray observations (e.g. \citealt{Mackay2010,Wright2011}), so comparing \gaia-RVS spectra with Chandra sources \citep{Evans2010} could lend further insight.   

It would also be beneficial to repeat this exercise with high resolution spectra (e.g. \citealt{bedell18}, \citealt{Brewer18}) to see what residuals at a greater resolution can tell us including intrinsic information in abundance measurements. This will let us go directly to the realm of understanding and quantifying how the data themselves are correlated \citep{Feeney2021}. We examine this idea of predicting data from data to get intuition for the variability around the Ca-triplet region in Appendix \ref{sec:neighbors}. 

The ability to study features like the cores and wings of the Ca-triplet with such a big data survey will help fill in the discrepancies between stellar models and data and help us answer fundamental questions in stellar astrophysics. This work is only further evidence that the \gaia\ mission has been revolutionary in our understanding of not only stellar astrophysics but also galaxy evolution, meeting at the unique intersection of galactic archaeology. The future release of \gaia-RVS spectra poses exciting prospects for this field with room to continue exploring the bridge between stars and the Galaxy's formation, evolution, and current state.

\acknowledgments
We thank Adam Wheeler, Chris Carr, Kirsten Blancato, and Rose Gibson for helpful discussions. 

R.R.~gratefully acknowledges the support of the Columbia University Bridge to the Ph.D.~Program in STEM.

%

\vspace{5mm}
\facilities{ARGOS, \gaia, RAVE, APOGEE}


\software{astropy \citep{2013A&A...558A..33A,astropyii}, sklearn \citep{sklearn}, scipy \citep{scipy}, Cannon \citep{cannon}
          }

%

\bibliographystyle{aasjournal}


\appendix \label{sec:append}
\renewcommand{\thesubsection}{\Alph{subsection}}
\counterwithin{figure}{subsection}
\counterwithin{table}{subsection}

\subsection{First 15 principal components for solar-metallicity stars}\label{sec:15pca}
\setcounter{figure}{0}
\begin{figure*}[htp]
    \centering
    \includegraphics[width=\textwidth,height=\textheight]{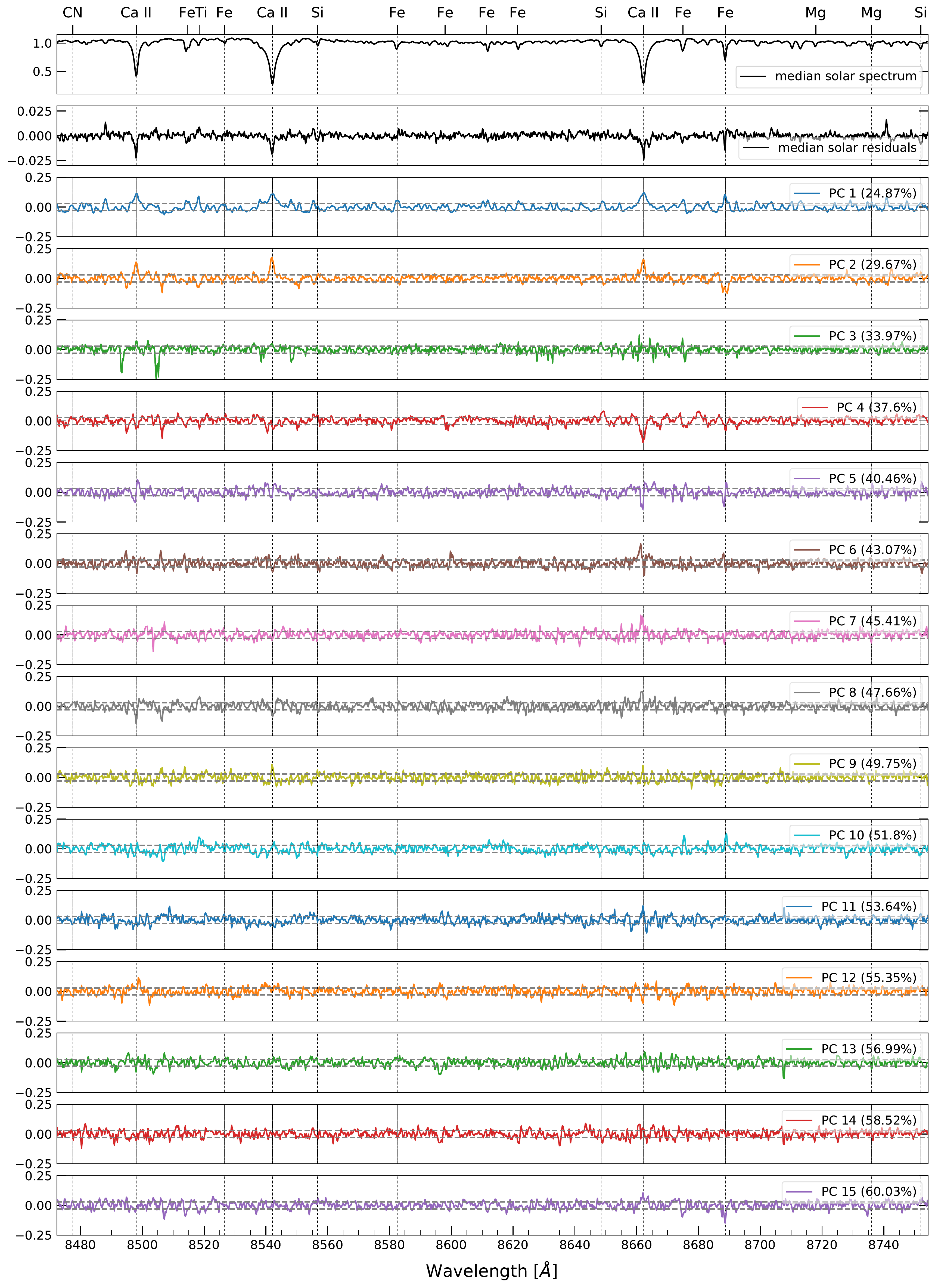}
    \caption{First 15 PCs for solar metallicity stars in sample with cumulative percentage of variance captured noted and PCA ''noise'' thresholds shown with grey dashed lines. \textit{Top Panel}: Median spectrum for solar-metallicity stars. \textit{Second Panel}: Median residuals for solar-metallicity stars. After PC 5, the subsequent components signal are so small, we deem them to be modeling noise.}
    \label{fig:allpcs}
\end{figure*}
\subsection{Labels v. Residuals at Ca Core}\label{sec:caresiduals}
\setcounter{figure}{0}
\begin{figure*}[htp]
    \centering
    \includegraphics[width=\textwidth]{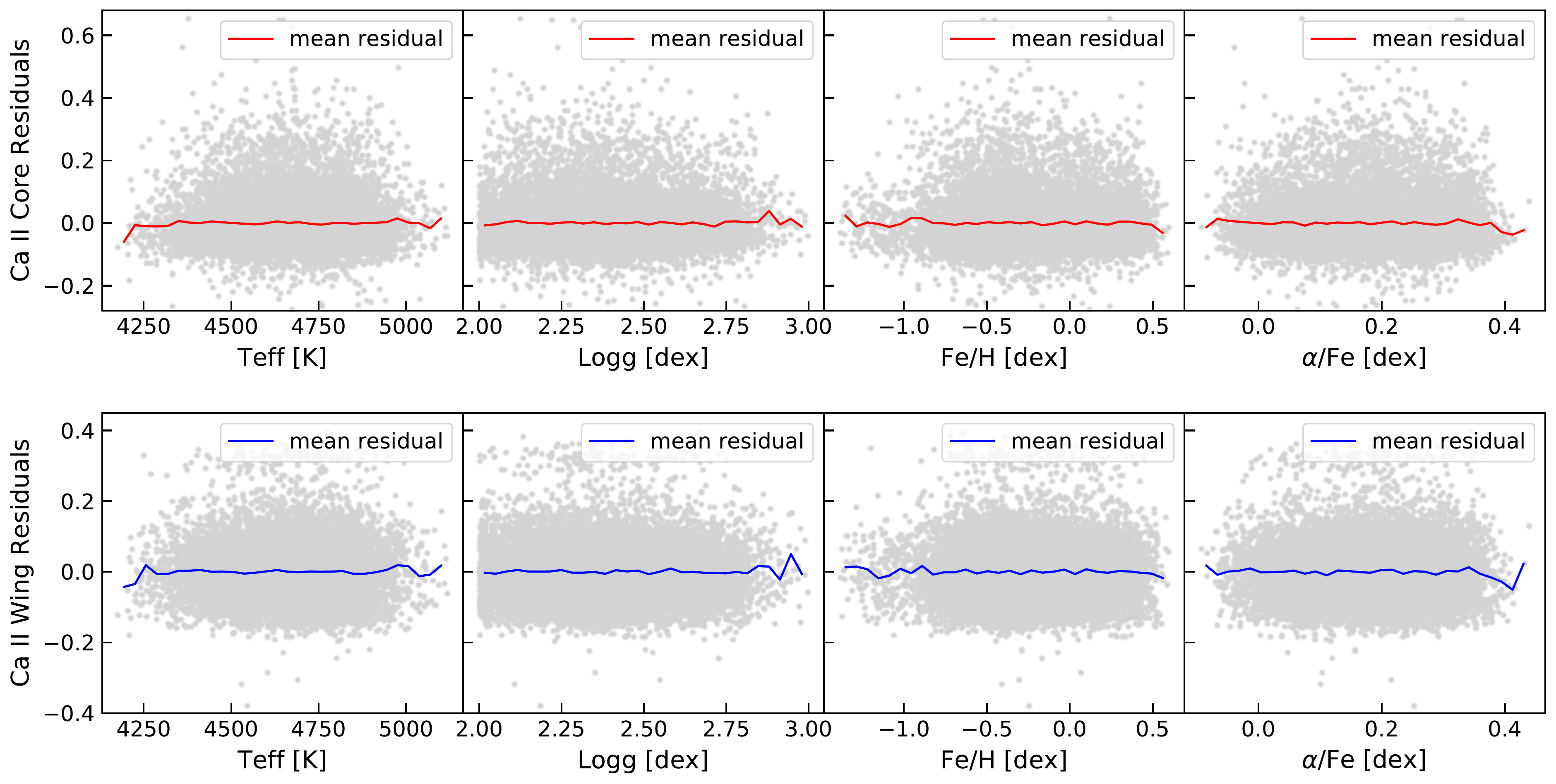}
   \caption{Labels versus Ca-triplet residual values. The relations are flat, indicating the polynomial models based on these label have modeled away any correlation. \textit{Top}: Second Ca-triplet core (8542.2 \AA) residual values versus \teff, \logg, \feh, and \alphafe. \textit{Bottom}: Second Ca-triplet wing (8547.3 \AA) residual values versus \teff, \logg, \feh, and \alphafe. }
    \label{fig:corewing}
\end{figure*}

\subsection{The variance in the Ca-triplet in nearest neighbourhood joint spectra}\label{sec:neighbors}

\setcounter{figure}{0}
\begin{figure*}
 \centering
\includegraphics[scale=0.235]{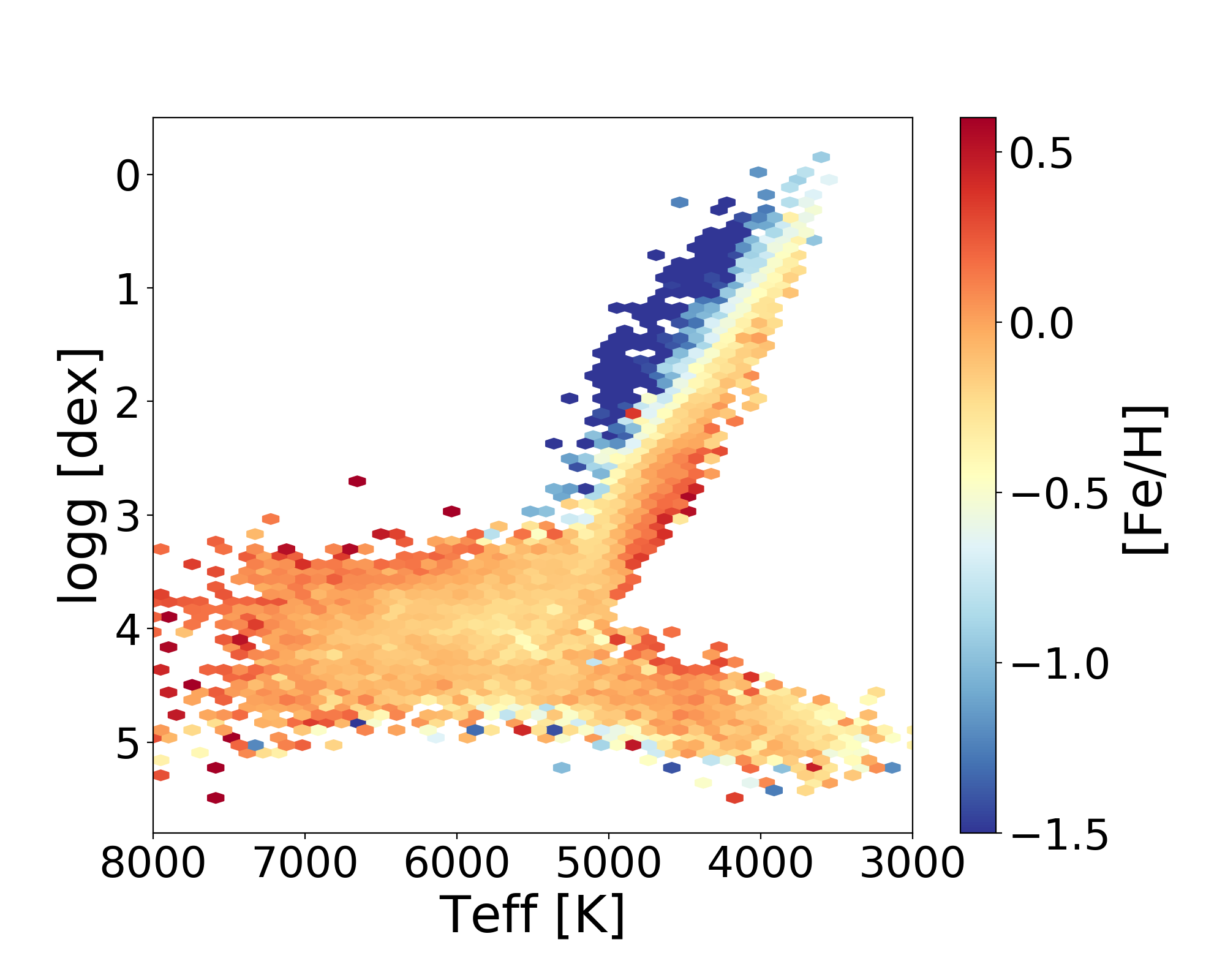}
\includegraphics[scale=0.235]{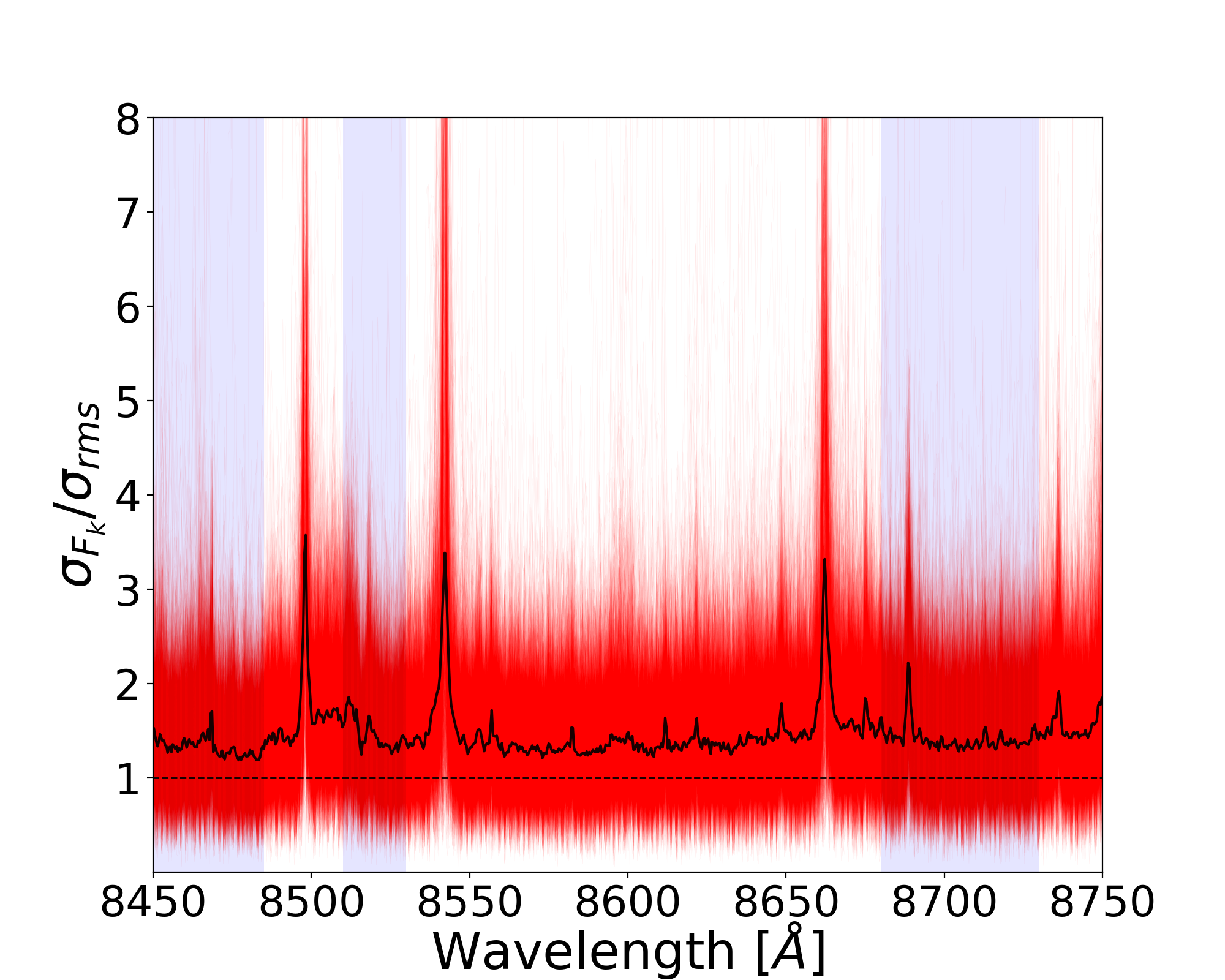}
\includegraphics[scale=0.235]{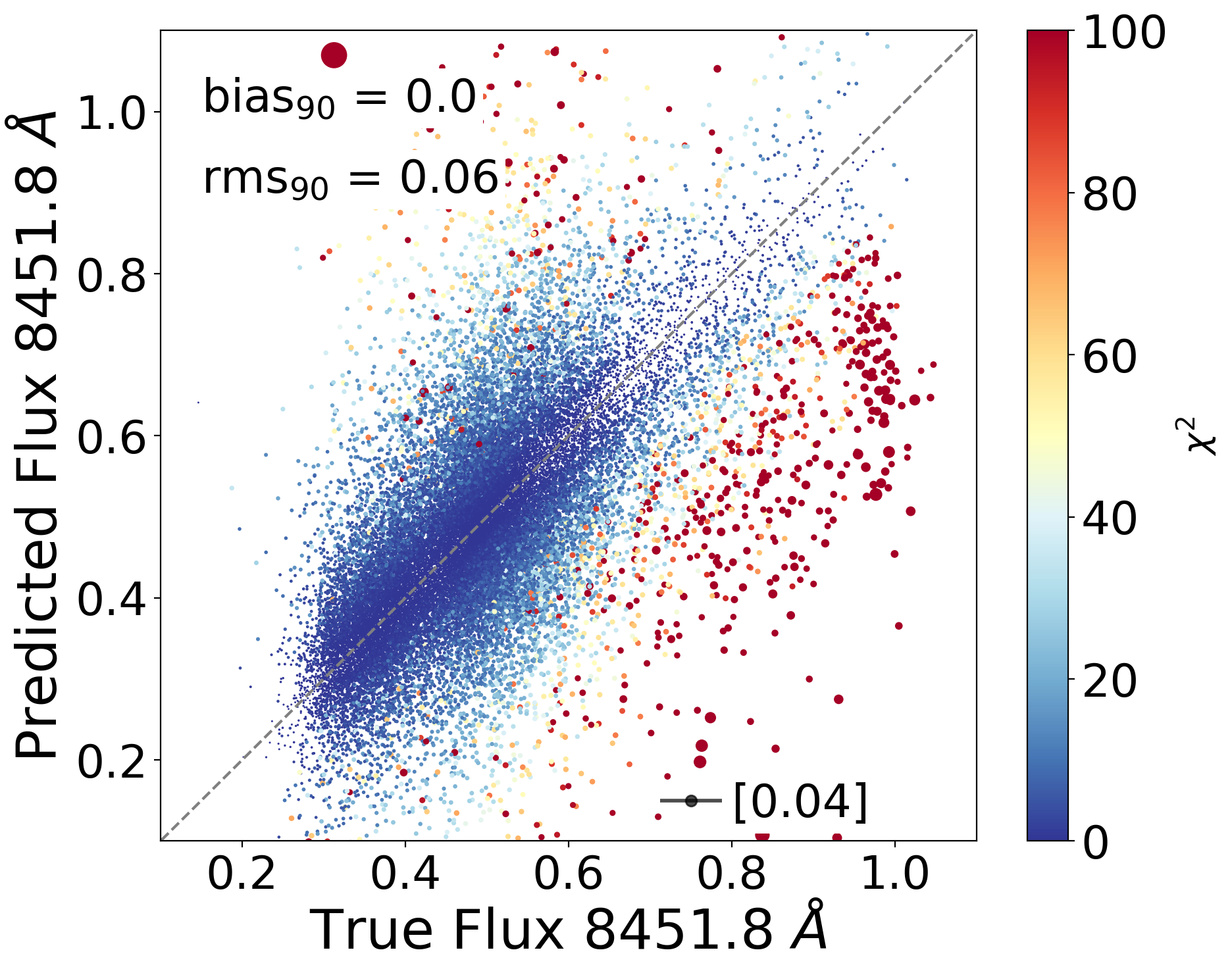}

      \caption{\textit{Left}: the \teff-\logg\ plane colored by \feh\ of the $\approx$ 50,000 RAVE test stars. \textit{Middle}: Ratio of the 1-$\sigma$ standard deviation of the k=5 nearest neighbours (defined using the blue windowed regions only) of each spectra to the quadrature sum of the error of the k=5 neighbours: 1000 sample individual stars are shown in red and the median in black. \textit{Right}: Regression model built from 1500 reference objects for a simple quadratic model of the flux in the windowed regions in the middle plot to predict the Ca-triplet core flux at 8541.8 \AA, shown for 40,000 test stars.}
    \label{fig:chispec}
\end{figure*}

With large data sets, it is possible to use partial data to model other partial data, by building a predictive model. Where the model's prediction fails can point to interesting objects. A systematic application of such an outlier detection approach is tangent space projection \citep{Wheeler2020}, which is promising to apply to \gaia-RVS spectra.

In Figure \ref{fig:chispec} we examine the RAVE spectra, to quantify the degree to which the Ca-triplet features vary relative to other regions of the spectra. Specifically, we generate nearest neighborhoods of spectra. We take stars across the \teff-\logg\ plane, shown in the left hand window of Figure \ref{fig:chispec}. We designate a reference window of spectra for the stars, shown in the blue region in the middle panel of Figure \ref{fig:chispec}, and a test region of spectra, the remaining spectra.  Using the 50,000 RAVE stars, across the \teff-\logg\ plane shown in the left panel of Figure \ref{fig:chispec} for every object we select the k nearest neighbour spectra, using an rms comparison of flux in the blue windowed regions shown in the middle panel of Figure \ref{fig:chispec}. Then for every nearest neighbourhood of k spectra we create an average spectra. The y-axis of the middle panel of Figure \ref{fig:chispec} shows, for 1000 joint spectra (averaged nearest neighbours), the ratio of the 1-$\sigma$ dispersion of their flux to the quadrature sum of their errors (both at each wavelength). This is to assess the relative variance of the joint spectra with respect to the total uncertainty, across each wavelength. A value of k = 5 nearest neighbours, for the full sample of N=50,000 RAVE test spectra return a level of $\approx$ 1 in the windowed regions used to determine the k nearest neighbours. (The smaller value of k the more identical the neighbours). Note that the middle of the spectra is not used to build the nearest neighbourhood for each star, yet is also has a ratio of variance to error that is similar to the windowed regions. The Ca-triplet regions, however, show a much larger variance than the error compared to the reference window, on average a factor of around $\approx$ 4, where the average of these individual examples is shown in black. This is indicative of the larger variance of this line flux in a neighbourhood, relative to the flux uncertainty, at fixed evolutionary state compared to the other spectral regions. This additional variance has no correlation with \teff, \logg, \feh, or \mg.

Finally, in the right hand panel of Figure \ref{fig:chispec} we do a simple regression predicting the single value of the Ca-triplet core flux at 8451 \AA\ using the windowed regions of flux in the middle panel. We use for this test $\approx$ 50,000 stars across the \teff-\logg\ plane. The reference set for training is of 1500 objects only with S/N $>$ 1000 and the test set is 40,000 objects with S/N $>$ 20. These points are coloured by the $\chi^2$ of the data minus the model (how well we predict relative to the error on the single flux point, where a high $\chi^2$ is indicate the prediction is poor relative to the uncertainty on the flux). The outliers around the 1:1 line in the right hand panel of Figure \ref{fig:chispec} with a high $\chi^2$ which have (upon inspection) significantly anomalous spectra compared to the rest of the sample. About 90 \% of the stars have an rms scatter of $\approx$ 0.06 dex, which is higher than the mean uncertainty of the flux (0.04 dex). The takeaways from this figure are twofold; (i) across the full sample, other spectral regions are not fully predictive of the flux at the Ca-triplet core, but the residual is small (the variation subtle) (ii) there are groups of stars where the true core flux deviates significantly from the prediction. Inspection of these stars is demonstrative their flux is significantly different from the majority of stars in the sample. The first point (i) likely speaks to dynamical activity in the star and the second (ii) populations with anomalous properties like binaries and emission stars. 

 This result using the spectra is consistent with what we see in Section \ref{sec:PCA}, where by the PCA components show systematic residuals around the Ca-triplet core flux, above the level of the flux uncertainty. With large data sets, unusual objects can be identified by failures in simple models where one part of the spectra is predicted by another \citep[e.g.][]{Wheeler2020, Kemp2018, Casey2019}. This indicates that \gaia-RVS spectra will be a rich source of data for empirically characterising the imprint of the stellar chromosphere and photosphere in the cores and also the wings of the triplet region, given the intrinsic information in this region not captured by correlations in the other spectral regions.



\end{document}